\documentclass[aps,prl,twocolumn,groupedaddress,showpacs,floatfix]{revtex4-1}

\usepackage{graphicx}
\usepackage[sort&compress]{natbib}

\begin{document}

\def\be{\begin{equation}}
\def\ee{\end{equation}}
\def\ba{\begin{eqnarray}}
\def\ea{\end{eqnarray}}
\def\vecr{{\bf r}}
\def\vecv{{\bf v}}

\title{Frequency Shifts in NIST Cs Primary Frequency Standards\\
 Due To Transverse RF Field Gradients}


\author{Neil Ashby}
\email[]{ashby@boulder.nist.gov}
\affiliation{National Institute of Standards and Technology}
\author{Stephan Barlow}
\email[]{sbarlow@boulder.nist.gov}
\affiliation{National Institute of Standards and Technology}
\author{Thomas Heavner}
\email[]{heavner@boulder.nist.gov}
\affiliation{National Institute of Standards and Technology}
\author{Steven Jefferts}
\email[]{jefferts@boulder.nist.gov}
\affiliation{National Institute of Standards and Technology}


\date{\today}

\begin{abstract}
A single-particle Green's function (propagator) is introduced to study the deflection of laser-cooled Cesium atoms in an atomic fountain due to RF field gradients in the Ramsey TE$_{011}$ cavity.  The deflection results in a state-dependent loss of atoms at apertures in the physics package, resulting in a frequency bias.   A model accounting only for motion in one dimension transverse to the symmetry axis of the fountain is discussed in detail and then generalized to two transverse dimensions.  Results for fractional frequency shifts due to transverse field gradients are computed for NIST F-1 and F-2 Cesium fountains.  The shifts are found to be negligible except in cases of higher RF power applied to the cavities.
\end{abstract}


\maketitle


\section{1. Introduction}
Frequency shifts in atomic clocks are of fundamental importance in the accuracy determination of the SI second, which is presently realized using laser-cooled Cs fountains operated by many standards laboratories around the world\cite{wynands05}.  The potential systematic bias due to momentum-changing interactions between the microwave interrogation field and the atoms undergoing Ramsey excitation have long been a source of concern, investigation, and conjecture.  Bord\`e and Wolf estimated the fractional frequency shift due to microwave recoil in atomic standards to be on the order of $\frac{\delta f}{f} \approx 10^{-16}$ \cite{wolfborde2004}. 

Recently, Gibble\cite{li11}\cite{gibble06} published a theory reinvestigating the microwave recoil shift along lines originally used by Cook\cite{cook78}\cite{cook87}.   Reference \cite{gibble06} uses Cook's methods in the optical domain to quantify the state-dependent deflection in the atomic trajectories due to gradients in the microwave field and the resulting frequency bias.  This work contains several unphysical results and predicts a frequency shift of order $10^{-16}$\cite{gibble06}; several Primary Frequency Standard Groups (NPL, PTB, SYRTE) are correcting for this bias.   In light of problematic results in \cite{gibble06} (e.g., the nonvanishing nature of the shift in the absence of microwave excitation), here we present a new theoretical treatment that extends the work by Cook or Gibble.  

The operation of the NIST fountains has been described in \cite{heavner05}. Figure 1 shows a simplified configuration of the NIST-F1 Cesium fountain.   A cloud of atoms is collected and laser-cooled to  $ \approx 0.5 \mu$K in optical molasses, then launched upwards through a state-selection region that results in a sample of atoms in  the $\left<3,0\vert\right.$ state.  The atoms pass into a Ramsey cavity (a cylindrical TE$_{011}$ cavity), where they are subject to a $\pi/2$ pulse of resonant RF radiation of frequency $f_0=9.192631770$ GHz that puts the atoms into a superposition of the two clock states $\left<4,0\vert\right.$ and $\left<3,0\vert\right.$.  The atoms then coast upwards into a drift region and fall back down through the cavity where they are subjected to a second $\pi/2$ pulse.  This causes some of  the atoms to make transitions to the upper hyperfine state.  The atoms then fall through a detection region which measures the numbers of atoms in each state; from the detected atom numbers in the two hyperfine states a relative transition probability is measured. 

\vbox to 24pt{}
\begin{figure}[ht]
\centering\includegraphics[width=2.0 truein]{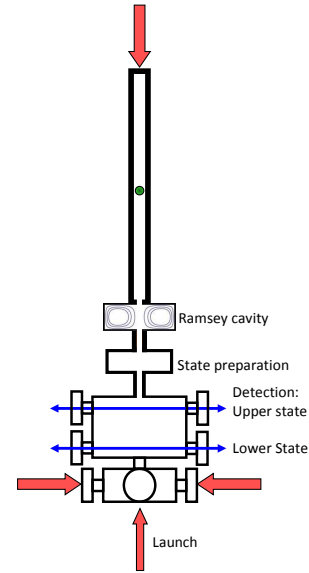}
\caption{Schematic diagram of the main components of NIST F-2.  See text for discussion.}
\end{figure}

Here we use a full wave-packet description of the atoms undergoing Ramsey excitation along with a full time-dependent solution to the Schrödinger equation that propagates the wave-packets through the two microwave interactions up through the detection step.  We find frequency shifts of order $10^{-17}$ in our atomic fountains\cite{heavner05} with our theory. (Because the microwave cavity is essentially identical in the two fountains at NIST and the geometry of the standards is similar, the results given here are representative of those for both fountains.)

The organization of this paper is as follows.  In Sect. 2 we introduce the Schr\"odinger equation for a two-state model of the hyperfine states of $^{133}$Cs, coupled to the axial component of the RF magnetic field in a cylindrically symmetric cavity.  The Hamiltonian of this problem includes kinetic energy, internal energy of the hyperfine states, and magnetic interaction energy of the spins with the magnetic field.  The RF field is assumed to be almost exactly resonant, a slight detuning is modeled by allowing the phase of the RF field to be different when the atoms pass through the cavity a second time.   In the presence of transverse microwave magnetic field gradients, which are dictated by vanishing of the transverse magnetic field at the cylindrical cavity boundaries, the transverse motion of the atoms is slightly affected.  This is described by the introduction of a propagator that exactly solves the Schr\"odinger equations after they are decoupled by a series of exact transformations.  We consider mainly the case of a $\pi/2$ pulse applied in each cavity but discuss up to $7\pi/2$ pulses in a later section.  The propagator accounts for the transverse recoil of the atoms due to the interaction of the spins with the field gradients.  Time development of the phases of the spinor components are discussed in Sect. 3.  Sections 4, 5, and 6 discuss construction of atom wavepackets and their quantum mechanical spreading during propagation through the apparatus.  In Section 7 a prototype one-dimension model of passage of a ball of laser-cooled atoms at temperature $T\approx 1 \mu$K, with spatial extent $\sigma_n$ in the transverse dimension, is developed in detail and solutions of the Schr\"odinger equation for balls entering the detector are presented.  Sect. 8 discusses the method of detection and Sect. 9 describes results for the one-dimensional case.  Sect. 10 generalizes the model to the full two-dimensional case of transverse motion.  The results for this case are described in Sect. 11. 

\section{2. Equations of Motion}
In this section we derive decoupled equations of motion for the atoms in the fountain and introduce a propagator (Green's function) that solves the spatial equations of motion for the atoms in a transverse magnetic field gradient.   The $\left<F,m_F\vert\right.=\left<4,0\vert\right.$ and $\left<F,m_F\vert\right.=\left<3,0\vert\right.$ hyperfine states of $^{133}$Cs, of intrinsic energies $\hbar \omega_a$ and  $\hbar \omega_b$, respectively, are coupled only by the $z-$component of an applied microwave magnetic field.\cite{scully89}    We consider a two-state model in which the energy separation of the hyperfine states is $\hbar \Delta=\hbar(\omega_a-\omega_b)$ and the zero of energy is halfway in between $\omega_a$ and $\omega_b$.  $\psi_a$ and $\psi_b$ denote the wavefunctions of atoms in the upper and lower hyperfine states, respectively.  In the presence of an applied RF field the equations of motion are:
\begin{eqnarray}\label{eq1}
i\hbar\frac{\partial\psi_a}{\partial t}=-\frac{\hbar^2}{2m}\nabla^2 \psi_a + \frac{\hbar \Delta}{2}\psi_a+\mu_{_B} g B_z \psi_b;\\
\label{eq2}
i\hbar\frac{\partial\psi_b}{\partial t}=-\frac{\hbar^2}{2m}\nabla^2 \psi_b - \frac{\hbar \Delta}{2}\psi_b+\mu_{_B} g B_z \psi_a.
\end{eqnarray}
For a TE$_{011}$ mode in a cylindrical cavity the applied high-frequency RF field may be represented by
\ba\label{eq3}
\mu_B g B_z=\hbar \pi b \cos(\omega t+\theta_1)\hbox to 1in{}\nonumber\\
=\hbar \pi b_0 \cos(\omega t+\theta_1)J_0(x_1 r/d)\sin(K z),
\ea
where $ K=(\omega^2/c^2-x_1^2/d^2)^{(1/2)}$, $b_0$ is a conveniently chosen measure of the amplitude, $x_1=3.83171...$ is the first zero of the ordinary Bessel function of order 1, $d$ is the cavity radius, and $c$ is the speed of light.  The applied field is assumed to be almost exactly on resonance:  $\omega  \approx  \Delta$.

In NIST F-1 and F-2, the symmetry axis of the cylindrical cavity is along the $z$-direction.  After  entering the cavity at the reference level $z=0$, atoms are subject to a half-sine wave pulse of RF energy (the $\sin(Kz)$ in Eq. (\ref{eq3})), coast upwards to height $h$, then fall back down through the cavity where another pulse is applied, then fall into a detector.  We allow the phases of the RF fields during cavity passage to be different: $\theta_1$ and $\theta_2$, respectively.  Time of passage through the cavities is denoted by $\tau$, time in the drift region by $T$, and time $T_d$ to fall from the bottom of the cavity into the detector.

The cavity has entry and exit apertures of radius $r_a<d/2$ and an atom can enter the cavity at some off-axis position ${x_c,y_c}$ where we think of this position as the center of a gaussian wave packet whose spread is small compared to the scale of distance over which the microwave magnetic field changes in the transverse direction.  A one-dimensional gaussian wave packet initially centered at $x_c$ can be constructed by superposing plane waves:
\begin{eqnarray}\label{packet}
\phi(x,t)=\sqrt{\frac{\sigma}{2\pi^{3/2}}}\int_{-\infty}^{\infty}dk e^{i(k(x-x_c)-\frac{\hbar k^2 t}{2m} -\sigma^2(k-k_0)^2/2}\nonumber\\
= \frac{1}{\pi^{1/4}}\frac{e^{i(k_0(x-x_c)-\hbar k_0^2 t/(2m))}}{\sqrt{\sigma +i \hbar t/(\sigma m)}}e^{-\frac{(x-x_c-\hbar k_0 t/m)^2}{2(\sigma^2+i\hbar t/m)}}.\hbox to .2in{}
\end{eqnarray}
The packet (\ref{packet}) is normalized to unity and satisfies the Schr\"odinger equation for a free particle,
\begin{equation}
i\hbar \frac{\partial \phi(x,t)}{\partial t}=-\frac{\hbar^2}{2m}\nabla^2 \phi(x,t).
\end{equation}
Quantum mechanical packet spreading occurs due to the terms proportional to $\hbar t/m$ in the denominators of (\ref{packet}).

It will be seen that all of the integrations performed as we follow the trajectory of an atom involve gaussian exponentials; such integrals can be performed in any order and the packets are normalized by means of the weighting functions in (\ref{packet}) so it will be convenient to simply assume that the packet entering the first cavity can be represented by a plane wave at $t=0$:
\begin{equation}
e^{i(k_x(x-x_c)+k_y(y-y_c))}.
\end{equation}

At an  off-axis position ${x_c,y_c}$ there is a transverse field gradient arising from the Bessel function in (\ref{eq3}).  Expanding in a Taylor series about such a position,
\begin{eqnarray}
\hbar \pi b_0 J_0(x_1 r/d)=\hbar \pi b_0 J_0(x_1\sqrt{x_c^2+y_c^2}/d)\nonumber\\
+2\gamma_x(x-x_c)+2\gamma_y(y-y_c),
\end{eqnarray}
where for example,
\begin{equation}\label{gammax}
\gamma_x=\frac{\hbar \pi b_0}{2}\frac{\partial J_0(x_1 r/d)}{\partial x}\Bigr|_{x_c,y_c}
\end{equation}

These gradients exert transverse forces on the spins and result in transverse displacements which are, however, quite tiny as will be seen.  That the gradients occur as a sum of terms in $x-x_c$ and $y-y_c$ in the equations of motion, makes it possible to solve (\ref{eq1}) and (\ref{eq2}) by separating variables.  

The time dependence in the equations of motion is simplified by passing to the ``rotating phase approximation."  We introduce a transformation of functions by means of
\begin{equation}\label{trantoD}
\left[
\begin{array}{c}
 \psi_{a}\\
 \psi_{b}
\end{array}
\right]=
\left[
\begin{array}{c}
        e^{-i(\omega t/2+\theta_1/2)}D_a \\ 
	e^{i(\omega t/2+\theta_1/2)}D_b
\end{array}
\right].
\end{equation}

Then introducing the exponential form for the $\cos(\omega t+\theta_1)$ appearing in (\ref{eq3}) and neglecting terms that oscillate with twice the hyperfine resonance frequency the equations of motion become
\begin{eqnarray}\label{eqsforD}
i\hbar\frac{\partial D_a}{\partial t}=-\frac{\hbar^2}{2m}\nabla^2 D_a +\frac{\hbar \pi b}{2} D_b+\frac{\hbar(\Delta-\omega)}{2}D_a;\nonumber\\
i\hbar\frac{\partial D_b}{\partial t}=-\frac{\hbar^2}{2m}\nabla^2 D_b -\frac{\hbar \pi b}{2} D_a+\frac{\hbar(\Delta-\omega)}{2}D_b.
\end{eqnarray}
We shall neglect the detuning terms in Eqs. (\ref{eqsforD}) above.  In the case where no spatial dependence is considered, it can be shown\cite{shirley01} that these terms cause a very small change in the width of the central Ramsey fringe; detuning does not itself cause a frequency shift.  We shall however keep the exponential phase terms in (\ref{trantoD}) as they play an important role in the discussion of detection.  The equations can then be decoupled by introducing the following linear combinations of wavefunctions:
\begin{equation}
f_+=\frac{1}{\sqrt{2}}(D_a+D_b);\hbox to .2in{} f_-=\frac{1}{\sqrt{2}}(D_a-D_b).
\end{equation}
The equations of motion become:
\begin{equation}\label{eq12}
i\hbar\frac{\partial f_{\pm}}{\partial t}=-\frac{\hbar^2}{2m}\nabla^2 f_{\pm} \pm \frac{\hbar \pi b}{2} f_{\pm}.
\end{equation}
Such linear combinations were introduced by Cook\cite{cook78} to treat motion of electric dipoles under the influence of a laser beam; the $f_+$ and $f_-$ packets are displaced in opposite directions, but consist at all times of equal contributions from the $a$ and $b$ spinor states, but with different phases.  The solution for $f_{-}$ can be obtained from the $f_+$ solution by changing the sign of $b$, so we shall consider only the $+$ sign and for convenience will drop the subscript.   

Launched atoms entering the cavity aperture arrive with a thermal distribution of velocities and hence with a distribution of arrival times; also the atom cloud may be clipped by the aperture edges.  Passage of the cloud through the cavity may thus entail a distribution of positions within the aperture as well as a distribution of times $\tau$ required to pass through the cavity.  We shall consider a particular packet centered at $(x_c,y_c)$ with a velocity $\hbar {\bf k}/m$ on entry to the cavity.  Transition probabilities for such an atom will be averaged over position and velocity at the end of the calculation.   As the packet traverses the cavity, it samples the half-sine wave dependence of the resonant RF field; this field in effect becomes a slowly varying time-dependent field because the packet will be well-localized compared to the cavity length.  The packet falls back through the cavity some time later; we assume that the second time of passage is the same as that during the first passage.  We therefore seek a reasonable approximation that allows description of the motion in the $z-$direction to be separated off; effects of field gradients in the $z-$direction cancel out and are not of interest in the present paper.  Therefore we assume
\begin{equation}\label{eq13}
f(x,y,z,t)=g(x,y,t)\phi(z,t);
\end{equation}
where $\phi(z,t)$ is of the general form (\ref{packet}).  Then (\ref{eq12}) becomes
\begin{equation}\label{eq14}
\phi(z,t)i\hbar\frac{\partial g}{\partial t}=-\frac{\hbar^2}{2m}\phi(z,t)\bigg(\frac{\partial^2g}{\partial x^2}+\frac{\partial^2g}{\partial y^2}  \bigg)+\frac{\hbar \pi b}{2}\phi(z,t)g.
\end{equation}
If we multiply by $\phi(z,t)^*$ the $z-$dependence can be simplified since integrating over all $z$,
\begin{eqnarray}\label{eq15}
\int \vert \phi(z,t) \vert^2 \sin(K z)\, dz=\sin\bigg(\frac{K \hbar k_{z0} t}{m}\bigg)\nonumber\\
\times e^{-\frac{K^2\hbar^2k_{z0}^2 t^2}{m^2}\bigg(\sigma^2+\hbar^2 t^2/(\sigma^2 m^2)\bigg)}.
\end{eqnarray}
The quantity appearing in the exponent in (\ref{eq15}) is extremely small during cavity passage and cannot affect the resonant frequency, so we shall neglect it and retain the time-dependent sine function.  Eq. (\ref{eq15}) then becomes
\begin{eqnarray}\label{eq16}
i\hbar \frac{\partial g}{\partial t}=-\frac{\hbar^2}{2m}\bigg(\frac{\partial^2g}{\partial x^2}+\frac{\partial^2g}{\partial y^2}  \bigg)\nonumber\hbox to .5in{}\\
+\sin(\kappa t)\bigg(\frac{\hbar \pi b_1}{2} +
\gamma_x(x-x_c)+\gamma_y(y-y_c)\bigg)g,
\end{eqnarray}
where $b_1= (\hbar \pi b_0/2) J_0(x_1\sqrt{x_c^2+y_c^2}/d)$, $\gamma_x$ is given in Eq. (\ref{gammax}) and where  $\kappa=K \hbar k_0/m$.   The first time-dependent term in (\ref{eq16}) term can be eliminated by letting\cite{shirley01}:
\begin{equation}
g(x,y,t)=e^{-ia(t)}h(x,y,t).
\end{equation}
Then if
\begin{equation}\label{eq18}
a(t)=\frac{ \pi b_1}{2}\int_0^t \sin{\kappa t'} dt', 
\end{equation}
Eq. (\ref{eq16}) reduces to
\begin{eqnarray}\label{effham}
i\hbar \frac{\partial h}{\partial t}=-\frac{\hbar^2}{2m}\bigg(\frac{\partial^2h}{\partial x^2}+\frac{\partial^2h}{\partial y^2}  \bigg)\nonumber\hbox to .5in{}\\
+\sin(\kappa t)\bigg( \gamma_x(x-x_c)+\gamma_y(y-y_c)\bigg)h.
\end{eqnarray}
The effective Hamiltonian on the right of (\ref{effham}) is a sum of terms that permit a solution by separation of variables into a product of factors of similar form.  If we let $h(x,y,t)=\alpha(x,t)\beta(y,t)$, then a solution of (\ref{effham}) is found if $\alpha$ satisfies
\begin{equation}\label{onedim}
i\hbar \frac{\partial\alpha}{\partial t}=-\frac{\hbar^2}{2m}\frac{\partial^2 \alpha}{\partial x^2}+\gamma_x (x-x_c)\sin(\kappa t)\alpha,
\end{equation} 
with an equation of motion of similar form for $\beta(y,t)$.  (There will be another set of solutions with opposite signs for $a(t)$ and $\gamma_x$, corresponding to selecting the ``-" option in (\ref{eq13}).)  We shall drop the subscript on $\gamma_x$ as long as we are discussing a one-dimensional model.

Separation of variables usually involves a ``separation constant;" in the present case this quantity becomes a function of time.  However its effect can be shown to cancel out of the product $\alpha(x,t)\beta(y,t)$.

At the boundary $z=0$, we take $t=0$ and require that the wavefunction be a plane wave.  Then (\ref{onedim}) can be solved with the propagator (Green's function)
\begin{equation}\label{G}
G_{\gamma}(x,t;x',0)=\sqrt{\frac{m}{2\pi i \hbar t}}e^{i S/\hbar},
\end{equation}
where
\begin{eqnarray}\label{eq22}
S=\frac{m}{2t}(x-x')^2- \hbox to 2.2in{}\nonumber\\
\frac{\gamma(x-x_c)}{\kappa}\bigg(-\cos(\kappa t)+\frac{\sin(\kappa t)}{\kappa t}\bigg)\hbox  to 1.35in{}\\
-\frac{\gamma(x'-x_c)}{\kappa}\bigg(1-\frac{\sin(\kappa t)}{\kappa t}\bigg)\hbox to 1.1in{}\nonumber\\
-\gamma^2\bigg( \frac{-2+2\kappa^2 t^2+2 \cos(2\kappa t)+\kappa t\sin(2 \kappa t)}{8\kappa^4 m t}\bigg).\nonumber
\end{eqnarray}
It is easily verified by straightforward calculation that,  
\begin{eqnarray}\label{eq23}
i\hbar\frac{\partial G_{\gamma}(x,t;x',0)}{\partial t}+\frac{\hbar^2}{2m}\frac{\partial^2G_{\gamma}(x,t;x',0)}{\partial x^2}\hbox to .9in{} \\
=i\hbar\delta(x-x')\delta(t)+\gamma (x-x_c) \sin(\kappa t)G_{\gamma}(x,t;x',0)\nonumber.
\end{eqnarray}
For a plane wave of the form $\psi(x,0)=e^{ik(x-x_c)}$ at $t=0$ entering the cavity, the solution of the Schr\"odinger equation in the cavity at time $t$ can be obtained from the propagator by integrating over the variable $x'$ at the initial time:
\begin{eqnarray}\label{psiincavity}
\psi_{\gamma}(x,t)=\int_{-\infty}^{\infty} dx' G(x,t;x',0)e^{ik(x'-x_c)} \hbox to .75in{}\nonumber\\
={\rm Exp}\bigg(-\frac{i\hbar k^2 t}{2m}+i\big(k-\frac{\gamma}{\hbar \kappa}(1-\cos(\kappa t)\big))(x-x_c) \nonumber\\
+\frac{ik\gamma t}{\kappa m}\big(1-\frac{\sin(\kappa t)}{\kappa t} \big)\hbox to 1.3in{}\\
-\frac{i\gamma^2}{8\hbar \kappa^3 m}\big(6\kappa t-8\sin(\kappa t)+\sin(2 \kappa t)  \big)\nonumber
\bigg);
\end{eqnarray}
a normalization constant would not be changed.  Similar propagators have been used by Scully, Schwinger, and Englert to describe the Stern-Gerlach effect in a static magnetic field,\cite{scully89},\cite{schwinger88},\cite{englert88}.  The propagator introduced in Eq. (\ref{G}) is one of a class of time dependent propagators that can be constructed by path-integral methods\cite{shulman65}.  A propagator for an electron in a static electric field gradient was first constructed by Kennard\cite{Kennard27}.   

In the present application, the transverse velocities of atoms are small compared with the launch velocity.  The launch velocity determines the total amount of time spent in the cavity by an atom.  The atoms are narrowly distributed about the launch velocity in the $z$ direction, so the values of the most likely $t$ that occurs in Eq. (\ref{psiincavity}) will be narrowly distributed about a value $\tau$ determined by the on-axis Ramsey pulse, $v_0 \tau =L$ where $L$ is the cavity length and $v_0$ is the central velocity at the entry aperture.  For a complete half-sine wave pulse on axis, $\kappa \tau=\pi$ and the exponentials in Eq. (\ref{psiincavity}), which are independent of $x$ and $x_c$, may be simplified to give:
\begin{eqnarray}\label{eq25}
\psi_{\gamma}(x,\tau)={\rm Exp}\bigg(-\frac{i\hbar k^2 \tau}{2m}+i\big(k-\frac{2\gamma \tau}{\hbar \pi}\big)(x-x_c) \nonumber\\
+\frac{ik\gamma \tau^2}{ m \pi} -\frac{i3\gamma^2 \tau^3}{4\hbar \pi^2 m} \big)
\bigg).\hbox to .5in{} 
\end{eqnarray}
Thus the propagator takes a plane wave into another plane wave at the entry into the drift region, with another phase determined by the kinetic energy of the particle, the wavenumber, and the field gradient, without changing the plane wave normalization.  The integrations over spatial variables can be performed by completing the squares in the exponents.  For example, to illustrate that order of integration is immaterial, a packet such as (\ref{packet}) is propagated through the cavity by calculating
\begin{eqnarray}\label{eq26}
\int dx' G_{\gamma}(x,\tau;x',0)\phi(x'-x_c,0)\hbox to .75in{}\nonumber\\
=\frac{1}{\pi^{1/4}}\frac{e^{i(k_0-2\gamma \tau/(\hbar \pi))(x-x_c)-i\hbar k_0^2 \tau/(2m)}}{\sqrt{\sigma+i\hbar t/(\sigma m)}}\hbox to .25in{}\nonumber\\
\times e^{-{(x-x_c-\hbar k_0 \tau/m+\gamma \tau^2/(m \pi))^2}/{(2(\sigma^2+i\hbar \tau/m))}}
\end{eqnarray} 
On the other hand if the packet is constructed at the end of the cavity using Eq. (\ref{eq26}), precisely the same result is obtained.  This justifies extending the range of integration over $x'$ to infinity, since a well-localized packet is small in size relative to the cavity aperture.

After exiting from the cavity, the wavefunction can be propagated to the end of the drift region with a free-particle propagator obtained from Eqs. (\ref{G}-\ref{eq22}) by setting $\gamma=0$. Thus, without constructing packets,
\ba\label{eq27}
\psi(x,\tau+T)\vert_{\gamma}=\int dx'G_{\gamma=0}(x,T;x',0)\psi_{\gamma}(x',\tau)\nonumber\\
=e^{-\frac{i\hbar k^2(\tau+T)}{2m}+i(k-\frac{2\gamma \tau}{\hbar \pi})(x-x_c)}\hbox to .7in{}  \nonumber\\
\times e^{+\frac{ik\gamma \tau(\tau+2T)}{m\pi}-\frac{i\gamma^2 \tau^2(3\tau+8T)}{4\hbar m \pi^2})}.\hbox to .6in{}
\ea
Such integrals may be performed with the help of a convergence factor\cite{Kennard27}; this is discussed in the Appendix.

Concatenation of these propagators can help in understanding the phase factors that enter the calculation.  For example, a propagator can be constructed that takes a particle from the entry aperture to the end of the drift region; thus
\begin{eqnarray}\label{propto2}
G_{free,\gamma,x_c}(x,\tau+T;x',0)\hbox to 1.5in{}\nonumber\\
=\int dx''G_{\gamma=0}(x,T;x'',0)G_{\gamma}(x'',\tau;x',0)\nonumber\\
=\sqrt{\frac{m}{2\pi i \hbar(\tau+T)}}e^{\frac{i m(x-x')^2}{2\hbar(\tau+T)}-\frac{i\gamma^2\tau^3(\tau+3T)}{4\hbar m\pi^2(\tau+T)}}\nonumber\\
\times e^{-\frac{i\gamma\tau(\tau(x-2x_c+x')+2T(x'-x_c))}{\hbar \pi(\tau+T)}}.\hbox to .3in{}
\end{eqnarray}

Propagation from the end of the drift region to the detector, through a cavity with a field gradient $\gamma_p$ with a packet centered at $x_p$ upon entry, can be calculated with a propagator similar to (\ref{propto2}):
\begin{equation}\label{eq29}
\centering{G_{free,\gamma_p,x_p}(x,\tau+T_d;x',0).}
\end{equation}
Then concatenating these two propagators will take an incoming plane wave all the way to the detector:
\begin{eqnarray}\label{eq30}
G_{final}(x,2\tau+T+T_d;x',0)=\hbox to 1.in{}\nonumber\\
\int dx''G_{free,\gamma_p,x_p}(x,\tau+T_d;x'',0)\hbox to 1in{}\nonumber\\
\times G_{free,\gamma,x_c}(x'',\tau+T,x',0)\hbox to .75in{}\nonumber\\
=\sqrt{\frac{m}{2\pi i \hbar T_t}}e^{\frac{im(x-x')^2}{2\hbar(T_t)}-\frac{i\gamma \gamma_p \tau^3(\tau+2T_d)}{\hbar m \pi^2T_t}}\hbox to .9in{}\nonumber\\
\times e^{-\frac{i\gamma^2\tau^3(4\tau+3T+3T_d)}{4\hbar m\pi^2T_t}-\frac{i\gamma_p^2\tau^2(4\tau^2+3\tau T+11 \tau T_d +8TT_d)}{4\hbar m \pi^2T_t}}\hbox to .3in{}\nonumber\\
\times e^{-\frac{i\gamma \tau(-2(T+T_d)(x_c-x')+\tau(x-4x_c+3x'))}{\hbar \pi T_t}}\hbox to .8in{}\nonumber\\
\times e^{-\frac{i\gamma_p \tau(2T(x-x_p)+\tau(3x-4x_p+x')+2T_d(-x_p+x'))}{\hbar \pi T_t}}.\hbox to .5in{}
\end{eqnarray}
where
\be\label{eq31}
T_t=2\tau+T+T_d.
\ee
Then if we insert the initial plane wave (centered at $x_c$), at the detector we find the plane wave
\ba\label{planetodetector}
\int dx''G_{final}(x,2\tau+T+T_d;x'',0)e^{ik(x"-x_c)}\nonumber\\
=e^{-\frac{i\hbar k^2(T_t)}{2m}+i(k-\frac{2\gamma \tau}{\hbar \pi})(x-x_c)-\frac{2i\gamma_p \tau(x-x_p)}{\hbar \pi}}\hbox to .2in{}\nonumber\\
\times e^{\frac{ik\gamma_p\tau(\tau+2T_d)}{m \pi}+\frac{ik\gamma \tau(3\tau+2T+2T_d)}{m\pi}-\frac{2i\gamma\gamma_p\tau^2(\tau+2T_d)}{\hbar m \pi^2}}\hbox to .1in{}\nonumber\\
\times e^{-\frac{i\gamma^2\tau^2(11\tau+8T+8T_d)}{4\hbar m\pi^2}-\frac{i\gamma_p^2\tau^2(3\tau+8T_d)}{4\hbar m\pi^2}}.\hbox to .6in{}
\ea

Of course this is only part of the solution since it only accounts for the dynamical particle motion. The internal states of the spinors will be treated in the next section.  To compress the equations we write the exponential in the result of Eq.  (\ref{planetodetector}) as
\be\label{dynamicalphase}
e^{i\Phi_f(\gamma,\gamma_p)}.
\ee

The function (\ref{planetodetector}) is an eigenstate of the momentum operator, since operating on (\ref{planetodetector}) with the momentum operator $i\hbar \partial/\partial x$, the momentum at the detector is
\be
\hbar k -\frac{2\gamma \tau}{m}-\frac{2\gamma_p\tau}{m}.
\ee
Other components of the wavefunction will differ in the signs of the contributions from $\gamma$ and $\gamma_p$.  Plane waves remain plane; the wavefronts acquire no curvature.  Thus there is no focusing of these solutions.   This is a consequence of the linear approximations for the field gradients in the dynamical equations of motion.  We next consider the development of the internal phases of the spinors.

\section{3.  Boundary Conditions; Spinor Phases}

In this section we discuss boundary conditions appropriate for the spinor part of the $f_{\pm}$ functions, given initial prepraration of the wavefunctions in an arbitrary superposition of $a$ and $b$ states.  In this section we consider here only the phase development of the spinors due to their internal energy; the dynamical phases have been treated above.  This discussion also applies to boundary conditions on the wavefunctions at the beginning of the second cavity passage.  

Suppose that upon first entry into the cavity the atomic spinor is in an arbitrary superposition of hyperfine states:
\begin{equation}\label{eq35}
\left[
\begin{array}{c}
 e^{-i\theta_1/2}D_a\\
 e^{i \theta_1/2}D_b
\end{array}
\right]\Bigg|_{t=0}=
\left[
\begin{array}{c}
        u_{a0} \\
       u_{b0} 
\end{array}
\right].
\end{equation}
Solving for the amplitudes $D_a$ and $D_b$ (Eq. (\ref{trantoD})) and substituting into Eq. (\ref{eq12}) gives initial conditions for $f_{\pm}$:
\begin{equation}\label{bccavity1}
f_{\pm}(0)=\frac{1}{\sqrt{2}}\big(u_{a0}e^{i\theta_1/2}\pm u_{b0}e^{-i\theta_1/2}\big)e^{ik(x'-x_c)},
\end{equation}
where $x'$ has been inserted in place of $x$ in anticipation of an integration.  If we were to assume the atoms are prepared in the lower hyperfine state before entering the first cavity, then $u_{a0}=0$ and $u_{b0}=1$ so the boundary condition (\ref{bccavity1}) would become 
\begin{equation}\label{eq37}
f_{\pm}(0)=\pm\frac{1}{\sqrt{2}}\big(e^{-i\theta_1/2}\big)e^{ik(x'-x_c)}.
\end{equation}

Combining (\ref{trantoD}) and (\ref{eq37}) in the general case, at the end of the first cavity passage the spinor functions are:
\ba\label{eq38}
u_a(\tau)=\frac{e^{-i(\omega \tau/2+\theta_1/2+a(\tau))}}{\sqrt{2}}f_+(0)\hbox to .3in{}\nonumber\\
+\frac{e^{-i(\omega \tau/2+\theta_1/2-a(\tau))}}{\sqrt{2}}f_{-}(0);
\hbox to .1in{}
\end{eqnarray}
\begin{eqnarray}\label{eq39}
u_b(\tau)=\frac{e^{i(\omega \tau/2+\theta_1/2-a(\tau))}}{\sqrt{2}}f_+(0)\nonumber\hbox to .3in{}\\
-\frac{e^{i(\omega\tau/2+\theta_1/2+a(\tau))}}{\sqrt{2}}f_{-}(0).
\hbox to .1in{}
\ea  
In the drift region the spinor states acquire additional phase factors due to their internal energy.  These phase factors are respectively
\begin{equation}\label{eq40}
e^{\mp i \Delta T/2}.
\end{equation}
Thus the spinor wavefunctions at the end of the drift region can be expressed as
\begin{eqnarray}\label{itera}
u_a(\tau+T)=\frac{e^{-i(\omega\tau+\Delta T+\theta_1)/2-ia}}{2}\big(u_{a0}e^{i\theta_1/2}+u_{b0}e^{-i\theta_1/2}\big)\nonumber\\
	+\frac{e^{-i(\omega\tau+\Delta T+\theta_1)/2+ia}}{2}\big(u_{a0}e^{i\theta_1/2}-u_{b0}e^{-i\theta_1/2}\big);\hbox to .4in{}
\end{eqnarray}
\begin{eqnarray}\label{iterb}
u_b(\tau+T)=\frac{e^{i(\omega\tau+\Delta T+\theta_1)/2-ia}}{2}\big(u_{a0}e^{i\theta_1/2}+u_{b0}e^{-i\theta_1/2}\big)\hbox to .1in{}\nonumber\\
        -\frac{e^{i(\omega\tau+\Delta T+\theta_1)/2+ia}}{2}\big(u_{a0}e^{i\theta_1/2}-u_{b0}e^{-i\theta_1/2}\big).\hbox to .5in{}
\end{eqnarray}
The signs of $\gamma$ and $a(\tau)$ occur in a given term with opposite signs, thus at the end of the calculation the dynamical phase factors can be matched with the spinor phase factors.  

In Eq. (\ref{bccavity1}), linear combinations of the spinor wavefunctions are combined to give boundary conditions for the decoupled functions $f_{\pm}$.  Similarly, linear combinations of Eqs. (\ref{itera}) and (\ref{iterb}) give boundary conditions for solution of the decoupled wave equations for second cavity passage.  In the second cavity, the phase factor $\exp(\pm i \theta_1/2)$  is replaced by $\exp(\pm i \theta_2/2)$; the value of $\theta_2$ will be discussed in Sect. 8.  

In order to simplify further development, to form the needed combinations we depart from a general treatment and assume the spinors are prepared in the lower hyperfine state, so $u_{a0}=0$, $u_{b0}=1.$
We therefore need the following linear combinations:
\begin{eqnarray}\label{eq43}
u_a(\tau+T)e^{i\theta_2/2}+\phi_b(\tau+T)e^{-i\theta_2/2}\hbox to .9in{}\\
=(\cos\Theta)e^{-ia-i\theta_1/2}+(+i\sin\Theta)e^{-ia-i\theta_1/2}, \nonumber\\
\label{eq44}
u_a(\tau+T)e^{i\theta_2/2}-\phi_b(\tau+T)e^{-i\theta_2/2}\hbox to .9in{}\\
=-(-i\sin\Theta)e^{ia-i\theta_1/2}-(-\cos\Theta)e^{ia-i\theta_1/2},\nonumber
\end{eqnarray}
where
\begin{equation}\label{eq45}
\Theta=\frac{1}{2}(\theta_2-\theta_1-\omega \tau-\Delta T ).
\end{equation}

The Ramsey fringe is determined by $\Theta$, which depends on the time $\tau$ through the first cavity and $T$ through the drift region.   

Eqs. (\ref{itera}) and (\ref{iterb}) give the spinors at the entry to the second cavity in terms of their initial values.  Equations of the same form must hold for the spinors at the detector in terms of their values at the entry to the second cavity.  Thus the spinors at the detector can be obtained by iterating (\ref{itera}) and (\ref{iterb}).  Such initial values have been given in Eqs. (\ref{eq43}-\ref{eq44}).  Therefore we can immediately write down the spinors at the detector.  The principal changes are: $\gamma$ and $a$ are replaced by $\gamma_p$ and $a_p$, respectively, $\theta_1$ is replaced by $\theta_2$, and $T$ is replaced by $T_d$.  Thus at the detector we have
\ba\label{spina}
u_a(2\tau+T+T_d)=\frac{e^{-i(\omega \tau+\Delta T_d)/2-i \theta_1/2-i \theta_2/2}}{2}\hbox to .6in{} \nonumber\\
\times\bigg( e^{-ia_p-ia} (\cos \Theta) 
+e^{ia_p-ia}(i\sin \Theta)\hbox to .75in{}\nonumber\\
+e^{-ia_p+ia}(-i\sin\Theta)
+e^{ia_p+ia}(-\cos \Theta)\bigg);\hbox to .5in{}
\ea
\ba\label{spinb}
u_b(2\tau+T+T_d)=\frac{e^{i (\omega \tau+\Delta T_d)/2-i\theta_1/2+i\theta_2/2}}{2}\hbox to 1in{} \nonumber\\
\times\bigg( e^{-ia_p-ia} (\cos \Theta)
+e^{ia_p-ia}(-i\sin \Theta)\hbox to 0.75in{}\nonumber\\
+e^{-ia_p+ia}(-i\sin\Theta)
-e^{ia_p+ia}(-\cos \Theta)\bigg).\hbox to .5in{}
\ea

The spinor components in (\ref{spina}) and (\ref{spinb}) are listed in the order $(\gamma,\gamma_p),(\gamma,-\gamma_p),(-\gamma,\gamma_p),(-\gamma, -\gamma_p)$ corresponding to $(a,a_p),(a,-a_p),(-a,a_p),(-a,-a_p)$. 

With the solution for the dynamical phase (\ref{dynamicalphase}), the wavefunctions at the detector can now be assembled:

\ba\label{fina}
u_a(2\tau+T+T_d)=\frac{e^{-i(\omega \tau+\Delta T_d)/2-i\theta_1/2-i\theta_2/2}}{2}\hbox to .25in{}\\
\times\bigg( e^{-ia_p-ia+i\Phi_f(\gamma,\gamma_p)} (\cos \Theta)\hbox to 1.in{} \nonumber\\
+e^{ia_p-ia+i\Phi_f(\gamma,-\gamma_p)}(i\sin \Theta)\hbox to 1.in{}\nonumber\\
+e^{-ia_p+ia+i\Phi_f(-\gamma,\gamma_p)}(-i\sin\Theta)\hbox to .75in{}\nonumber\\
+e^{ia_p+ia+i\Phi_f(-\gamma,-\gamma_p)}(-\cos \Theta)\bigg);\hbox to .5in{} \nonumber
\ea
\ba\label{finb}
u_b(2\tau+T+T_d)=\frac{e^{i(\omega \tau+\Delta T_d)/2-i\theta_1/2+i\theta_2/2}}{2}\hbox to .25in{} \\
\times\bigg( e^{-ia_p-ia+i\Phi_f(\gamma,\gamma_p)} (\cos \Theta)\hbox to 1in{} \nonumber\\
+e^{ia_p-ia+i\Phi_f(\gamma,-\gamma_p)}(-i\sin \Theta)\hbox to 1in{}\nonumber\\
+e^{-ia_p+ia+i\Phi_f(-\gamma,\gamma_p)}(-i\sin\Theta)\hbox to .75in{}\nonumber\\
+e^{ia_p+ia+i\Phi_f(-\gamma,-\gamma_p)}(+\cos \Theta)\bigg);\hbox to .5in{}\nonumber
\ea

	Eqs. (\ref{fina}) and (\ref{finb}) provide the complete solution for plane wave spinors passing through the apparatus.  In the next section we discuss the dynamical phases arising from transverse particle motion.
\begin{figure}[ht]
\centering\includegraphics[width=3.125 truein]{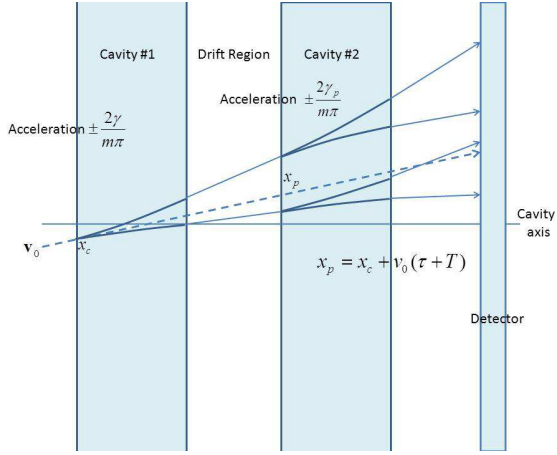}
\caption{Schematic illustration of trajectories of packets through the system; each cavity causes separation into two packets, each with equal numbers of $a$ and $b$ spinor states. The ordinate represents the direction transverse to the cavity axis.}
\end{figure}

\section{4. Construction of Wave Packets}

We construct wavepackets at the detector by multiplying by the weighting function $\exp(-\sigma^2 (k-k_0)^2/2)$, as in (\ref{packet}), and integrating over $k$.  Every term in (\ref{fina}) and (\ref{finb}) acquires a normalization factor
\be\label{eq50}
N_p=\frac{1}{\sqrt{\sqrt{\pi}\big(\sigma+\frac{i\hbar T_t}{\sigma m}}\big)}.
\ee
The expressions become quite cumbersome.  We shall illustrate the result in only one case, the first term in (\ref{fina}).  The terms in the exponent involving the wave vector $k$ are
\ba\label{eq51}
-\sigma^2 (k-k_0)^2/2 -\frac{i \hbar k^2T_t}{2m}+i k(x-x_c)\nonumber\\
+\frac{i k \gamma_p \tau (\tau+2T_d)}{m \pi}+\frac{i k\gamma \tau(3\tau+2T+2T_d)}{m \pi}.\hbox to .2in{}
\ea
This term then becomes
\be\label{eq52}
\frac{N_p}{2}e^{-ia_p-ia+i\Phi_{packet}(k_0,\gamma,\gamma_p)}(\cos\Theta)
\ee
where
\ba\label{firsttermphase}
\Phi_{packet}(k_0,\gamma,\gamma_p)=-\frac{\hbar k_0^2 T_t}{2m}-\frac{2\gamma \tau}{\hbar \pi}(x-x_c)\hbox to .4in{}\nonumber\\
-\frac{2\gamma_p \tau}{\hbar \pi}(x-x_p)-\frac{2\gamma\gamma_p\tau^2(\tau+2T_d)}{4\hbar m \pi^2}\hbox to .4in{}\\
+k_0\big(x-x_c+\frac{\gamma\tau(3\tau+2T+T_d)}{m\pi}+\frac{\gamma_p\tau(\tau+2T_d)}{m \pi} \big)\nonumber\\
+i\frac{(x-x_c-\frac{\hbar k_0T_t}{m}+\frac{\gamma \tau(3\tau+2T+3T_d)}{m \pi}+\frac{\gamma_p\tau(\tau+2T_d))}{m \pi})^2}{2(\sigma^2+i\hbar T_t/m)}.\nonumber
\ea
The quadratic term in $(x-x_c...)^2$ in (\ref{firsttermphase}) clearly shows where the packet is centered.  One can also backtrack and obtain the packet center at the exit aperture by setting $T_d=0$.

The packets for the hyperfine states at the detector are therefore:
\ba\label{finapacket}
\Psi_a(2\tau+T+T_d)\hbox to 2in{}\nonumber\\=
\frac{N_p}{2} 
\times\bigg( e^{-ia_p-ia+i\Phi_{packet}(k_0,\gamma,\gamma_p)} (\cos \Theta)\hbox to 1in{} \nonumber\\
+e^{ia_p-ia+i\Phi_{packet}(k_0,\gamma,-\gamma_p)}(i\sin \Theta)\hbox to 1in{}\nonumber\\
+e^{-ia_p+ia+i\Phi_{packet}(k_0,-\gamma,\gamma_p)}(-i\sin\Theta)\hbox to .75in{}\nonumber\\
+e^{ia_p+ia+i\Phi_{packet}(k_0,-\gamma,-\gamma_p)}(-\cos \Theta)\bigg);\hbox to .5in{}
\ea
\ba\label{finbpacket}
\Psi_b(2\tau+T+T_d)\hbox to 2in{}\nonumber\\=\frac{N_p}{2}
\times\bigg( e^{-ia_p-ia+i\Phi_{packet}(k_0,\gamma,\gamma_p)} (\cos \Theta)\hbox to 1in{} \nonumber\\
+e^{ia_p-ia+i\Phi_{packet}(k_0,\gamma,-\gamma_p)}(-i\sin \Theta)\hbox to 1in{}\nonumber\\
+e^{-ia_p+ia+i\Phi_{packet}(k_0,-\gamma,\gamma_p)}(-i\sin\Theta)\hbox to .75in{}\nonumber\\
+e^{ia_p+ia+i\Phi_{packet}(k_0,-\gamma,-\gamma_p)}(+\cos \Theta)\bigg).\hbox to .5in{}
\ea

To save writing, we shall refer to these terms in order as $\psi_{1a},\psi_{2a}...\psi_{3b},\psi_{4b}$ respectively.  The probability of finding a particle in the upper hyperfine state at the detector will be 
\be\label{sqr}
\big<\big|\Psi_a \big|^2\big>
\ee
where $<>$ means appropriately averaged over thermal velocities and integrated over the apertures.  

In computing probabilities such as in (\ref{sqr}), each of the squared terms gives 16 contributions.  The forms for the solutions given above allow us to cancel many terms even without knowing much about the functions $\Phi_{packet}$.  We are going to compress the notation by labeling the terms in the solutions with subscripts 1 through 4.  Thus for example, for the contribution from the square of the lower hyperfine state, the product of the first term, times the complex conjugate of the third term, will be denoted by
\ba\label{defofintegral}
ie^{-2ia}\cos\Theta \sin\Theta P_{13b}=N_p^2\hbox to 1.3in{}\nonumber\\
\times\int\big< (\cos \Theta e^{-ia-ia_p} e^{i\Phi_{packet}(\gamma,\gamma_p)})\hbox to .75in{}\nonumber\\
\times (i\sin\Theta e^{-ia+ia_p}e^{i\Phi_{packet}^*(-\gamma,\gamma_p)})\big>.\hbox to .5in{}
\ea
Then normalization of the wavefunction at the detector is computed by means of
\ba\label{inta}
\int\big<\big|\Psi_a \big|^2 \big>=\frac{1}{4}\bigg((\cos \Theta)^2(P_{11a}+P_{44a})\hbox to 1.1in{}\nonumber\\
-(\cos \Theta)^2(e^{-2ia_p-2ia}P_{14a}+e^{2ia_p+2ia}P_{41a})\hbox to .9in{}\nonumber\\
+(\sin \Theta)^2(P_{22a}+P_{33a})\hbox to 1.9in{}\nonumber\\
-(\sin \Theta)^2(e^{2ia_p-2ia}P_{23a}+e^{-2ia_p+2ia}P_{32a})\hbox to .7in{} \nonumber\\
+i\cos\Theta \sin\Theta \big(-e^{-2ia_p}P_{12a}+e^{2i a_p}P_{21a} \hbox to .7in{}\nonumber\\
+e^{-2ia}P_{13a}-e^{2ia}P_{31a}\hbox to .9in{}\nonumber\\
-e^{-2ia}P_{24a}+e^{2ia}P_{42a}\hbox to .9in{}\nonumber\\
+e^{-2ia_p}P_{34a}-e^{2ia_p}P_{43a}\big)\bigg)\hbox to .7in{}
\ea
and
\ba\label{intb} 
\int\big<\big|\Psi_b \big|^2 \big>=\frac{1}{4}\bigg((\cos \Theta)^2(P_{11b}+P_{44b})\hbox to 1.1in{}\nonumber\\
+(\cos \Theta)^2(e^{-2ia_p-2ia}P_{14b}+e^{2ia_p+2ia}P_{41b})\hbox to .9in{}\nonumber\\
+(\sin \Theta)^2(P_{22b}+P_{33b})\hbox to 1.9in{}\nonumber\\
+(\sin \Theta)^2(e^{2ia_p-2ia}P_{23b}+e^{-2ia_p+2ia}P_{32b})\hbox to .7in{} \nonumber\\
+i\cos\Theta \sin\Theta \big(e^{-2ia_p}P_{12b}-e^{2i a_p}P_{21b} \hbox to .7in{}\nonumber\\
+e^{-2ia}P_{13b}-e^{2ia}P_{31b}\hbox to .9in{}\nonumber\\
-e^{-2ia}P_{24b}+e^{2ia}P_{42b}\hbox to .9in{}\nonumber\\
-e^{-2ia_p}P_{34b}+e^{2ia_p}P_{43b}\big)\bigg)\hbox to .7in{}
\ea
These integrals have been defined so that when $``\int"$ in (\ref{inta}) or (\ref{intb}) is interpreted as an integration over all space, 
\be\label{symmetry1}
P_{iia}=P_{iib}=1,\quad i=1,2,3,4.
\ee
It can then be shown that for all $i$ and $j$,
\be\label{symmetry2}
P_{ija}=P_{ijb}.
\ee
The normalization condition reduces to
\ba
1=1+\frac{2}{4}i\cos\Theta\sin\Theta\hbox to 1.3in{}\nonumber\\
\times\int\big(e^{-2ia}P_{13a}-e^{2ia}P_{31a}\hbox to .5in{}\nonumber\\
-e^{-2ia}P_{24a}+e^{2ia}P_{42a} \big).
\ea
In can be shown that in addition to the above symmetry properties of the integrals, we have when integrating over all space
\be
P_{13a}=P_{24a};\quad P_{31a}=P_{24a}.
\ee
Therefore all terms proportional to $\cos\Theta\sin\Theta$ on the right side of the normalization condition cancel and the solution is correctly normalized.  In order for the integrals to converge, it is necessary to construct wavepackets as in (\ref{packet}) before squaring.
 
\section{5. Packet centers}

{\it ``It is the mark of an educated mind to rest satisfied with the degree of precision which the nature of the subject admits and not to seek exactness where only an approximation is possible.}--Aristotle
\vskip 10pt
The factors multiplying the terms in large parentheses in (\ref{fina}) and (\ref{finb}) cannot affect any probability computed from the squares of the wavefunctions; these factors will therefore be dropped.

Wave packets can be constructed at the beginning or end of the drift region by taking weighted superpositions of $e^{ik(x-x_c)}$ as in (\ref{packet}).  Figure 1 illustrates, greatly exaggerated, what happens to the wave packet trajectories.  The decoupled components travel in opposite directions in the first cavity, with an effective acceleration $\pm 2 \gamma /(m \pi)$. For a typical off-axis field gradient, the separation in position is only about a nanometer.  The velocity difference is about 15 nm/second, so at the end of the drift region if $T=1$ sec, the packet separation is only about 30 nm.  On the other hand a typical transverse velocity due to a finite temperature of $\approx .5\mu K$ is 0.008 m/sec. The displacements due to the thermal velocity distribution are $\approx 3 \times 10^5$ larger than the displacements due to transverse field gradients.  The latter are also small on the scale of changes in the transverse field gradients themselves.  We therefore use a value of $\gamma$, denoted by $\gamma_p$, in the second cavity that accounts for the change in the central position of the packet during time $\tau+T$.  A packet entering the first cavity with position $x_c$ and average velocity $\hbar k_{0x}/m$ will end up, on average, at the entry to the second cavity with position
\begin{equation}
x_p=x_c+\hbar k_{0x}(\tau+T)/m.
\end{equation}
The field gradient at this position will be denoted by
\begin{eqnarray}
\gamma_p=-\frac{\hbar \pi b x_1^2 x_p}{4 d^2}\hbox to 1.2in{}\nonumber\\
=-\bigg(\frac{\hbar \pi b x_1^2 x_c}{4 d^2}+\frac{\hbar^2 \pi b x_1^2 k_{0x}(\tau+T)}{4 d^2 m}\bigg),
\end{eqnarray}
and the second term will be accounted for when performing thermal averages.  On exiting the first cavity, the packet will be centered at
\be
x=x_c+\frac{\hbar k_{0x}\tau}{m}\mp\frac{\gamma \tau^2}{m\pi};
\ee
at the end of the drift region, the center will be at
\be
x=x_c +\frac{\hbar k_{0x}(\tau+T)}{m}\mp\frac{\gamma \tau(\tau+2T)}{m\pi};
\ee
after exiting from the second cavity, the center will be at
\be
x=x_c +\frac{\hbar k_{0x}(2\tau+T)}{m}\mp\frac{\gamma \tau(3\tau+2T)}{m\pi}\mp\frac{\gamma_p\tau^2}{m\pi};
\ee
and finally at the detector the center will be
\ba
x=x_c +\frac{\hbar k_{0x}(2\tau+T+T_d)}{m}\mp\frac{\gamma \tau(3\tau+2T+2T_d)}{m\pi}\nonumber\\
\mp\frac{\gamma_p\tau(\tau+2T_d)}{m\pi}\hbox to 1.2in{}.
\ea
The parameter $a_p(\tau)$ in the second cavity will also be affected by transversFe forces.   If the on-axis value of $a(\tau)$ is given by a normal $\pi/2$ pulse, then for an atom on the axis
\begin{equation}
a_{axis}=b \tau=\pi/4.
\end{equation}
At the entry to the first cavity we have
\begin{equation}\label{aentry}
a=a_{axis}\bigg(1-\frac{x_1^2x_c^2}{4 d^2}\bigg),
\end{equation}
and therefore at the entry to the second cavity
\begin{equation}\label{apentryto2}
a_p=a_{axis}\bigg(1-\frac{x_1^2(x_c+\hbar k_{0x}(\tau+T)/m)^2}{4 d^2}\bigg).
\end{equation}

\section{6. Launch conditions; Thermal Averaging}

In NIST F-2, atoms are collected, cooled, and launched from a trap some dozens of centimeters below the cavity entry aperture.  We let $z_L<0$, $v_L>0$ and $t_L<0$ be the launch position, velocity and time of launch, such that the atom clouds are centered at the entry aperture at $z=0,\ t=0$ with entry velocity $v_0=v_L+g t_L $.  Here $z$ is positive upwards and the acceleration of gravity is $-g$.  We treat the cloud of atoms as collisionless, characterized by an initial spread $\sigma_n$ and temperature $T_n$ in the transverse directions, and spread $\sigma_p$ and temperature $T_p$ in the z-dimension.  The atom distribution is described by a product of exponential distribution  functions of the following form:
\begin{equation}
f_n(x,v_x,t)=\frac{1}{2\pi \sigma_n}\sqrt{\frac{m}{k_B T_n}}e^{-\frac{(x-v_x(t+t_L))^2}{2 \sigma_n^2}}e^{-\frac{mv_x^2}{2k_BT_n}},
\end{equation}
\begin{eqnarray}
f_p(z,v_z,t)=\frac{1}{2\pi \sigma_p}\sqrt{\frac{m}{k_B T_p}} e^{
-\frac{m(v_z-v_L+g(t+t_L))^2}{2k_B T_p}}\nonumber\\
\times e^{-\frac{(z+d_L-v_z(t+t_L)-g(t+t_L)^2/2)^2}{2\sigma_p^2}}.
\end{eqnarray}
Then the complete distribution function is
\be\label{distfunct}
f(\vecr,\vecv,t)=f_n(x,v_x,t)f_n(y,v_y,t)f_p(z,v_z,t).
\ee
This distribution function satisfies the collisionless Boltzmann equation,
\be
\frac{\partial f}{\partial t}+\vecv \cdot \nabla f -g\frac{\partial f}{\partial v_z}=0.
\ee
At $t=t_L$, the cloud is centered at $z=z_L$.  In the transverse direction, the half-width of the cloud is $\sigma_n$ at $t=-t_L$ (in a single tranverse dimension) and spreads to $(\sigma_n^2+k_B (t+t_L)^2 T_n/m)^{1/2}$ after time $t$.

There are many contributions to the width of the wavepackets and to the atom balls; generally these combine in quadrature.    First, there is the initial half-width $\sigma$ at launch, which is not known precisely.  Due to quantum mechanical spreading of a packet, there is a second contribution to the half-width that increases with time but is inversely proportional to $\sigma$.  If $\sigma$ is small, quantum mechanical spreading will become large and contribute significantly to clipping.  The value of $\sigma$ has been set equal to the approximate thermal wavelength of a Cs$^{133}$ atom, about 200 nanometers.  In Figure 9 we illustrate the effect of changing this assumption, and verify that the results do not depend significantly on the choice of this parameter.  Third, there is a contribution arising from thermal averaging over the transverse velocities.  At temperatures $0.5 \mu K$ and higher, and after a Ramsey time in the neighborhood of 1 second, this contribution dominates the spreading of the ball.   There are numerous additional contributions to spreading, of the order of a nanometer or less, arising from transverse field gradients and off-axis positions of the atoms.  These contributions are automatically included in the calculations.

\section{7. One-Dimensional Model}

The detector measures the numbers of particles in the cloud in each of the hyperfine states and the following ratio is computed:
\be\label{observable}
\frac{\int\big<\vert \Psi_a \vert^2 \big>}{\int\big<\vert \Psi_a \vert^2 \big>+\int\big<\vert \Psi_b \vert^2 \big>},
\ee
where the integrals go over the initial aperture with a weight function determined by the Boltzmann distribution, after forming wave packets.  Thus we average with
\be
\int_{-D/2}^{D/2}f_n(x_c,v_x,0)dx_c dv_x \vert \Psi(x) \vert^2
\ee
where $D=.005$ meters is the aperture radius.  We also integrate over the exit aperture:
\be
\int\big<\vert \Psi_a \vert^2 \big>=\int_{-D/2}^{D/2}dx \int_{-D/2}^{D/2}f_n(x_c,v_x,0)dx_c dv_x \vert \Psi_a(x) \vert^2
\ee
with the understanding that the integrals now go over a finite region, we still denote the contributions with notations such as $P_{11a}$, etc.  The dependence of the spinor part of the wavefunctions is not affected by such restrictions. 

The numerator of (\ref{observable}) is of the form
\be
\int\big<\vert \Psi_a \vert^2 \big>=A_a+B_a\cos2\Theta+C_a \sin2\Theta,
\ee
where
\ba
A_a=\frac{1}{8}\bigg(P_{11a}+P_{22a}+P_{33a}+P_{44a}\nonumber\\
-e^{-2ia_p-2ia}P_{14a}-e^{2ia_p+2ia}P_{41a}\nonumber\\
-e^{2ia_p-2ia}P_{23a}-e^{-2ia_p+2ia}P_{32a}\bigg)
\ea
\ba
B_a=\frac{1}{8}\bigg(P_{11a}-P_{22a}-P_{33a}+P_{44a}\nonumber\\
-e^{-2ia_p-2ia}P_{14a}-e^{2ia_p+2ia}P_{41a}\nonumber\\
+e^{2ia_p-2ia}P_{23a}+e^{-2ia_p+2ia}P_{32a}\bigg)
\ea
\ba
C_a=\frac{i}{8}\bigg(-e^{-2ia_p}P_{12a}+e^{2ia_p}P_{21a}\hbox to .2in{}\nonumber\\
+e^{-2ia}P_{13a}-e^{2ia}P_{31a}\hbox to .4in{}\nonumber\\
-e^{-2ia}P_{24a}+e^{2ia}P_{42a}\hbox to .4in{}\nonumber\\
+e^{-2ia_p}P_{34a}-e^{2ia_p}P_{43a}\bigg).\hbox to .2in{}
\ea
Similarly for the other hyperfine state,
\ba
A_b=\frac{1}{8}\bigg(P_{11b}+P_{22b}+P_{33b}+P_{44b}\nonumber\\
+e^{-2ia_p-2ia}P_{14b}+e^{2ia_p+2ia}P_{41b}\nonumber\\
+e^{2ia_p-2ia}P_{23b}+e^{-2ia_p+2ia}P_{32b}\bigg)
\ea
\ba
B_b=\frac{1}{8}\bigg(P_{11b}-P_{22b}-P_{33b}+P_{44b}\nonumber\\
+e^{-2ia_p-2ia}P_{14b}+e^{2ia_p+2ia}P_{41b}\nonumber\\
-e^{2ia_p-2ia}P_{23b}-e^{-2ia_p+2ia}P_{32b}\bigg)
\ea
\ba
C_b=\frac{i}{8}\bigg(e^{-2ia_p}P_{12b}-e^{2ia_p}P_{21b}\hbox to .2in{}\nonumber\\
+e^{-2ia}P_{13b}-e^{2ia}P_{31b}\hbox to .4in{}\nonumber\\
-e^{-2ia}P_{24b}+e^{2ia}P_{42b}\hbox to .4in{}\nonumber\\
-e^{-2ia_p}P_{34b}+e^{2ia_p}P_{43b}\bigg).\hbox to .2in{}
\ea
The denominator of (\ref{observable}) is the sum of contributions from both hyperfine states and is therefore
\be
A+B \cos 2\Theta +C \sin 2\Theta
\ee
where 
\ba
A=A_a+A_b=\frac{1}{4}\big(P_{11a}+P_{22a}+P_{33a}+P_{44a} \big);\nonumber\\
B=B_a+B_b=\frac{1}{4}\big(P_{11a}-P_{22a}-P_{33a}+P_{44a} \big);\\
C=C_a+C_b=\frac{i}{4}\big(e^{-2ia}P_{13a}-e^{2ia}P_{31a}\nonumber\\
-e^{-2ia}P_{24a}+e^{2ia}P_{42b}\big).\hbox to 1in{}
\ea
and where we have made use of the relations (\ref{symmetry1}, \ref{symmetry2}).

\section{8. Detection}
	Let the numbers of particles detected in the $a$ and $b$ hyperfine states be
\ba
n_a=\big< \int \vert \Psi_a \vert^2 \big>;\\
n_b=\big< \int \vert \Psi_b \vert^2 \big>.
\ea
where it is understood that the spatial integrals only go over the apertures, and that averages over transverse velocity are performed with weight functions derived from (\ref{distfunct}).

The transition probabilities found in the preceding calculations depend on the phase difference $\theta_2-\theta_1$ only through the quantity
\begin{equation}\label{eq93}
2\Theta=\theta_2-\theta_1-(\omega \tau+\Delta T).
\end{equation}
At the instant the atom enters the cavity for the second time, the angle $\theta_2$ includes the advance of phase of the microwave field, as well as any additional phase angle $\alpha$ that is imposed on the RF field during the drift time.  Thus at resonance the phase of the microwave field satisfies   
\ba\label{Theta}
\theta_2=\theta_1+\omega(\tau+T)+\alpha;\nonumber\\
2\Theta=\alpha+(\omega-\Delta)T.\hbox to .2in{}
\ea
Due to the contibutions of the coefficients $C_a,C_b$ to the transition probabilities, the center of the line will be slightly shifted from its nominal value at $\omega=\Delta$.

In the F-2 fountain, the central Ramsey fringe is located, corresponding ideally to $\alpha=0,\ \omega=\Delta$.
The numbers of particles in the upper hyperfine state are measured part way down the line profile, on opposite sides of the line in successive balls; this corresponds to setting $(\omega-\Delta)T=\pm \pi/2$ or if $\omega=\Delta$,  $\alpha=\pm \pi/2$.  A servo locates the line center in such a way that the numbers observed on the two sides of the line are equal.  A frequency shift error $\delta \omega $ would appear through an additional term $\Delta \rightarrow\Delta+\delta \omega$ in (\ref{Theta}):
\begin{equation}\label{eq95}
2\Theta_{\pm}=\pm \pi/2-\delta \omega T.
\end{equation}
After thermal averages and averages over the aperture are taken, theory gives the following predictions for the observables:
\be\label{posside}
\frac{n_+}{n_++n_-}\bigg|_{+\alpha}=\frac{A_a+B_a \cos(2\Theta_+)+C_a \sin(2\Theta_+)}{A+B \cos(2\Theta_+)+C \sin(2\Theta_+)};
\ee
\be\label{negside}
\frac{n_+}{n_++n_-}\bigg|_{-\alpha}=\frac{A_a+B_a \cos(2\Theta_-)+C_a \sin(2\Theta_-)}{A+B \cos(2\Theta_-)+C \sin(2\Theta_-)}.
\ee
If the coefficients $C_a, C_b$ vanished, there could be no frequency shift since the transition probabilities would be symmetric with respect to reflection about the line center.

We consider two cases.  First, setting the ratios (\ref{posside}) and (\ref{negside}) equal, expanding for small $\delta \omega $ and solving we find that $\delta \omega $ is proportional to the factor 
\be
\cos(\alpha)(C B_a-B C_a)-(A C_a-C A_a).
\ee
If $\alpha$ were chosen to satisfy the above relation, the shift of the center of the line would be unobservable.  For NIST F-2, $\alpha \approx 1.7$ radians, whereas the value of $\alpha$ actually used in making measurements is $\pi/2=1.57$ radians.  Figure 3 shows the calculated shift as a function of the angle $\alpha$ for a reasonable set of parameters.  The measured shift decreases from the actual shift at $\alpha=0$ to a smaller value at $\alpha=\pm \pi/2$.  
\begin{figure}
\centering\includegraphics[width=3.5 truein] {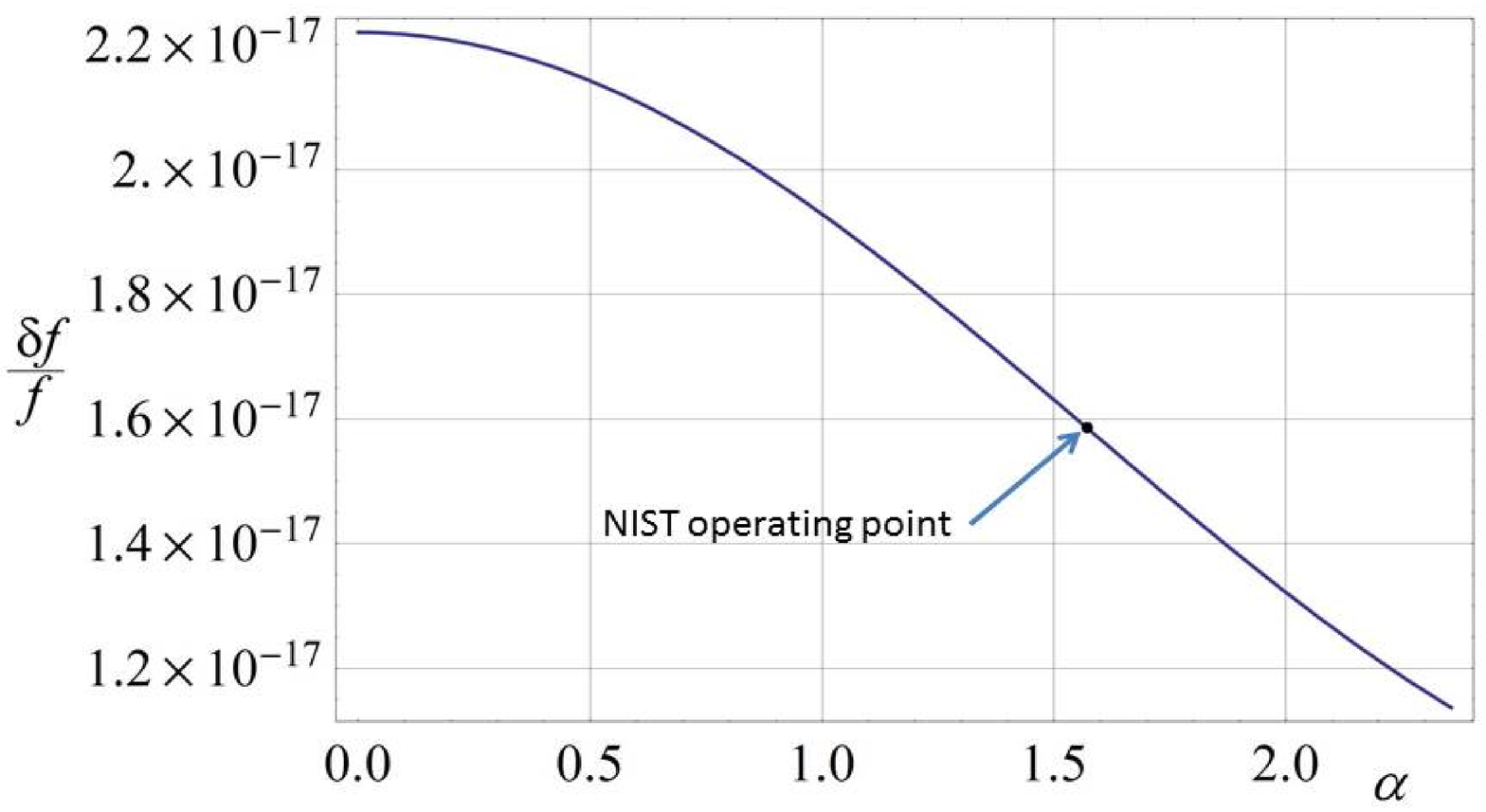}
\caption{Calculated fractional frequency shift as a function of the angle $\alpha$ in the two-dimensional case.  The values of toss height and temperature are 1.1 meters and 0.5 $\mu$K, respectively.}
\end{figure}

Second, for the actual operating conditions of NIST F-1 and F-2, $\alpha=\pm \pi/2$.  Solving for the shift as in the previous case,
\be
\delta \omega =\frac{A C_a-C A_a}{A B_a T}
\ee
the fractional frequency error is thus
\begin{equation}
\frac{\delta \omega}{\omega}=\frac{AC_a-C A_a}{2\pi \times (9.192\times 10^9)TB_aA}.
\end{equation}

\section{9. Results: One-Dimensional Case}
Figure 4 plots the fractional frequency shift as a function of toss height $h$, at a temperature of $0.5 \mu{\rm K}$.  The fractional frequency shift as a function of initial atom cloud temperature is plotted in Figure 5 for a toss height 0.75 meters.  
\begin{figure}
\centering\includegraphics[width=3.5 truein] {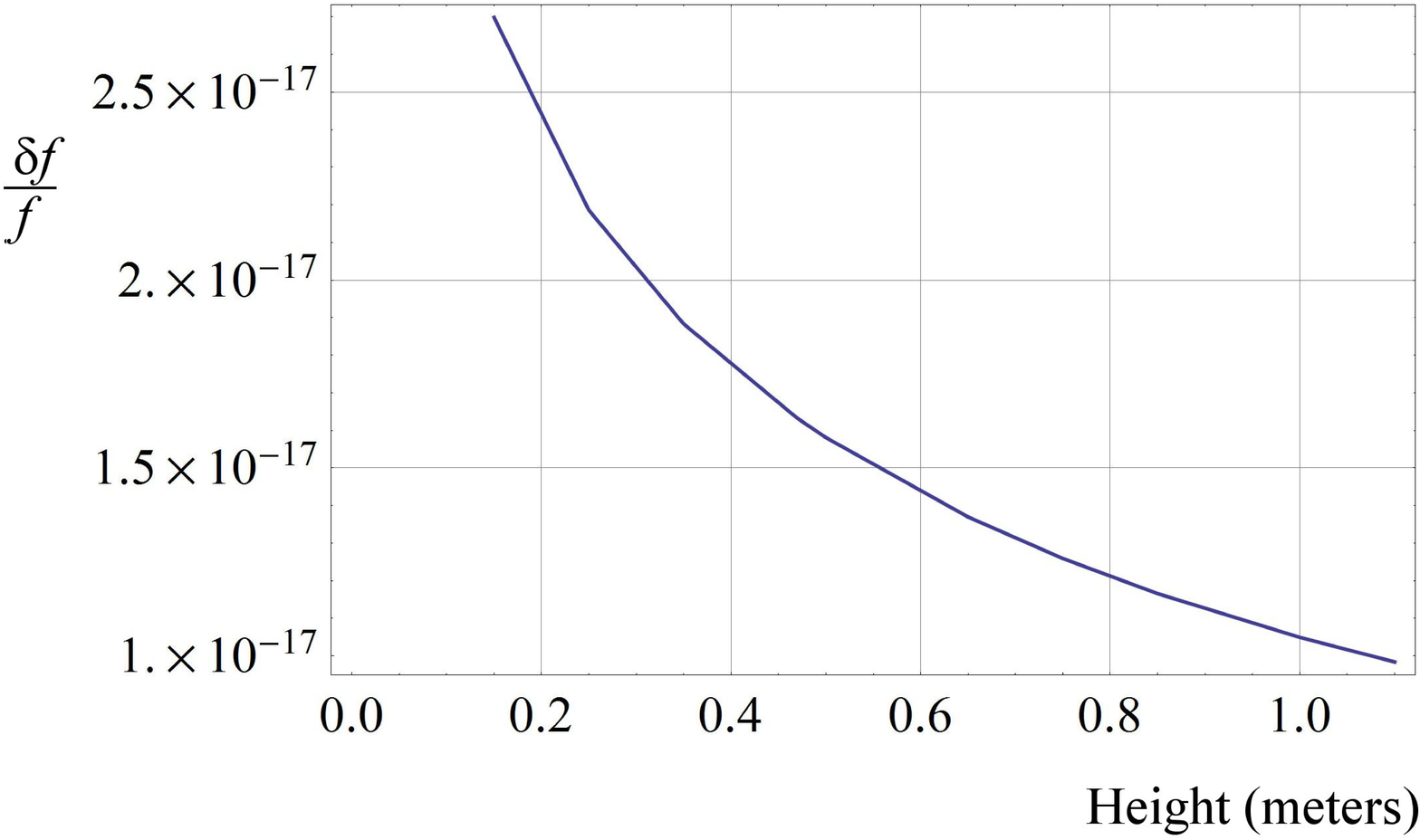}  
\caption{Fractional shift as a function of toss height, at temperature $0.5 \mu{\rm K}$; one-dimensional calculation}
\end{figure}
\begin{figure}
\centering\includegraphics[width=3.5 truein]{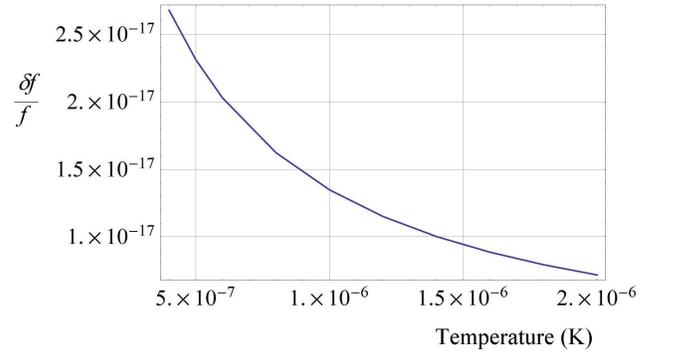}   
\caption{Fractional frequency shift for various atom cloud temperatures at a toss height 0.75 meters; one-dimensional calculation}
\end{figure}

\section{10. Two-dimensional model}

	The theory may be extended to two dimensions in a straightforward way.  Eq. (\ref{effham}) admits a solution that is a product $\alpha(x,t)\beta(y,t)$.  The function $\beta$ is formally identical to $\alpha$ with replacements $x\rightarrow y$, $x_c \rightarrow y_c$, $\gamma_x \rightarrow \gamma_y$, $x_p \rightarrow y_p$,  $\gamma_{xp} \rightarrow \gamma_{yp}$, $k_0=k_{0x}\rightarrow k_{0y}$, $\vert  N_p \vert^2 \rightarrow \vert N_p \vert^4 $.  Factors such as $e^{\pm a \pm a_p} \cos(\Theta)$ are formally unchanged, but $a$ and $a_p$ are evaluated at an off-axis point corresponding to $(x_c,y_c)$. Thus, Eqs. (\ref{aentry}) and (\ref{apentryto2}) become
\be\label{eq101}
a=a_{axis}\bigg(1-\frac{x_1^2(x_c^2+y_c^2)}{4d^2}\bigg);
\ee
\ba
a_p=a_{axis}\bigg( 1-\frac{x_1^2(x_c+\hbar k_{0x}(\tau+T)/m)^2}{4 d^2}\hbox to .25in{}\nonumber\\
-\frac{x_1^2(y_c+\hbar k_{0y}(\tau+T)/m)^2}{4 d^2}\bigg).\hbox to .1in{}
\ea
Cavity phase factors involving $a$ and $a_p$ are sums in the exponent, which becomes a product of exponential phase factors.   
The dynamical phase factors in Eqs. (\ref{finapacket}) and (\ref{finbpacket}) are augmented by factors of the form
\be
e^{i\Phi_{packet}(k_{0y},\pm \gamma_y,\pm \gamma_{yp})}.
\ee  
Let
\ba
\Phi_{pkt}(k_{0x},\gamma_x,\gamma_{xp},k_{0y},\gamma_y,\gamma_{yp})\hbox to 1.5in{}\\
=\Phi_{packet}(k_{0x},\gamma_x,\gamma_{xp})+\Phi_{packet}(k_{0y},\gamma_y,\gamma_{yp}).\nonumber
\ea
(We suppress the dependence on $x,x_c,x_p,y,y_c,y_p$ to save writing.)  Then the wavefunctions at the detector are 

\ba\label{finapacket2}
\Psi_a(2\tau+T+T_d)=\frac{N_p^2}{2}\hbox to 1.5in{}\\
\times\bigg( e^{-ia_p-ia+i\Phi_{pkt}(k_{0x},\gamma_x,\gamma_{xp},k_{0y},\gamma_y,\gamma_{yp})} (\cos \Theta)\hbox to .4in{} \nonumber\\
+e^{ia_p-ia+i\Phi_{pkt}(k_{0x},\gamma_x,-\gamma_{xp},k_{0y},\gamma_y,-\gamma_{yp})}(i\sin \Theta)\hbox to .2in{}\nonumber\\
+e^{-ia_p+ia+i\Phi_{pkt}(k_{0x},-\gamma,_x\gamma_{xp},k_{0y},-\gamma_y,\gamma_{yp})}(-i\sin\Theta)\hbox to .1in{}\nonumber\\
+e^{ia_p+ia+i\Phi_{pkt}(k_{0x},-\gamma_x,-\gamma_{xp},k_{0y},-\gamma_y,-\gamma_{yp})}(-\cos \Theta)\bigg);\hbox to .1in{}\nonumber
\ea
\ba\label{finbpacket2}
\Psi_b(2\tau+T+T_d)=\frac{N_p^2}{2}\hbox to 1.5in{}\\
\times\bigg( e^{-ia_p-ia+i\Phi_{pkt}(k_{0x},\gamma_x,\gamma_{xp},k_{0y},\gamma_y,\gamma_{yp})} (\cos \Theta)\hbox to .4in{} \nonumber\\
+e^{ia_p-ia+i\Phi_{pkt}(k_{0x},\gamma,_x,-\gamma_{xp},k_{0y},\gamma_y,-\gamma_{yp})}(-i\sin \Theta)\hbox to .2in{}\nonumber\\
+e^{-ia_p+ia+i\Phi_{pkt}(k_{0x},-\gamma,_x,\gamma_{xp},k_{0y},-\gamma_y,\gamma_{yp})}(-i\sin\Theta)\hbox to .1in{}\nonumber\\
+e^{ia_p+ia+i\Phi_{pkt}(k_{0x},-\gamma,_x,-\gamma_{xp},k_{0y},-\gamma_y,-\gamma_{yp})}(+\cos \Theta)\bigg).\hbox to .1in{}\nonumber
\ea
The Boltzmann distribution function for motion in the $y$ direction is formally the same as that for motion in the $x$ direction and the net particle distribution function that depends on $x$ and $y$ is just a product of two similar exponential functions.  Similarly, the quantum mechanical probability that depends on the two transverse coordinates $(x,y)$ is just a product of two functions of the same form.  If we let $P_{ija}$ or $P_{ijb}$ denote integrals over $x$ as in Eq. (\ref{defofintegral}), and $Q_{ija}$ or $Q_{ijb}$ denote corresponding integrals over $y$, then Eqs. (\ref{inta}-\ref{intb}) are valid when we make the replacements
\be
P_{ija} \rightarrow P_{ija}Q_{ija};\quad P_{ijb}\rightarrow P_{ijb}Q_{ijb}.
\ee
The discussion of detection is unchanged.

When numerically averaging over a circular entry aperture, $x$ and $y$ are restricted to
\be
(x^2+y^2)^{1/2}\le r_a; \quad (x_c^2+y_c^2)^{1/2}\le r_a .
\ee 
For thermal averaging over the initial velocity distributions, the transverse part of the Boltzmann distribution takes the following two-dimensional form:
\ba
f(x,v_x,y,v_y,0)=f_n(x,v_x,0)f_n(y,v_y,0)\nonumber\\
=\frac{m}{(2\pi\sigma_n)^2k_B T}e^{-\frac{m}{2k_BT_n}(v_x^2+v_y^2)}\nonumber\\
\times e^{-\frac{(x-v_xt_L)^2}{2\sigma_n^2}}e^{-\frac{(y-v_y t_L)^2}{2\sigma_n^2}}.
\ea

\section{ 11. Results: Two-Dimensional Model}

  In Figure 6 we plot the results for transmission through circular apertures of radius 5 mm for temperature $T=0.5\mu$K, as a function of toss height.  This should be compared with the one-dimensional results plotted in Figure 3.  
\begin{figure}
\centering\includegraphics[width=3.5 truein]{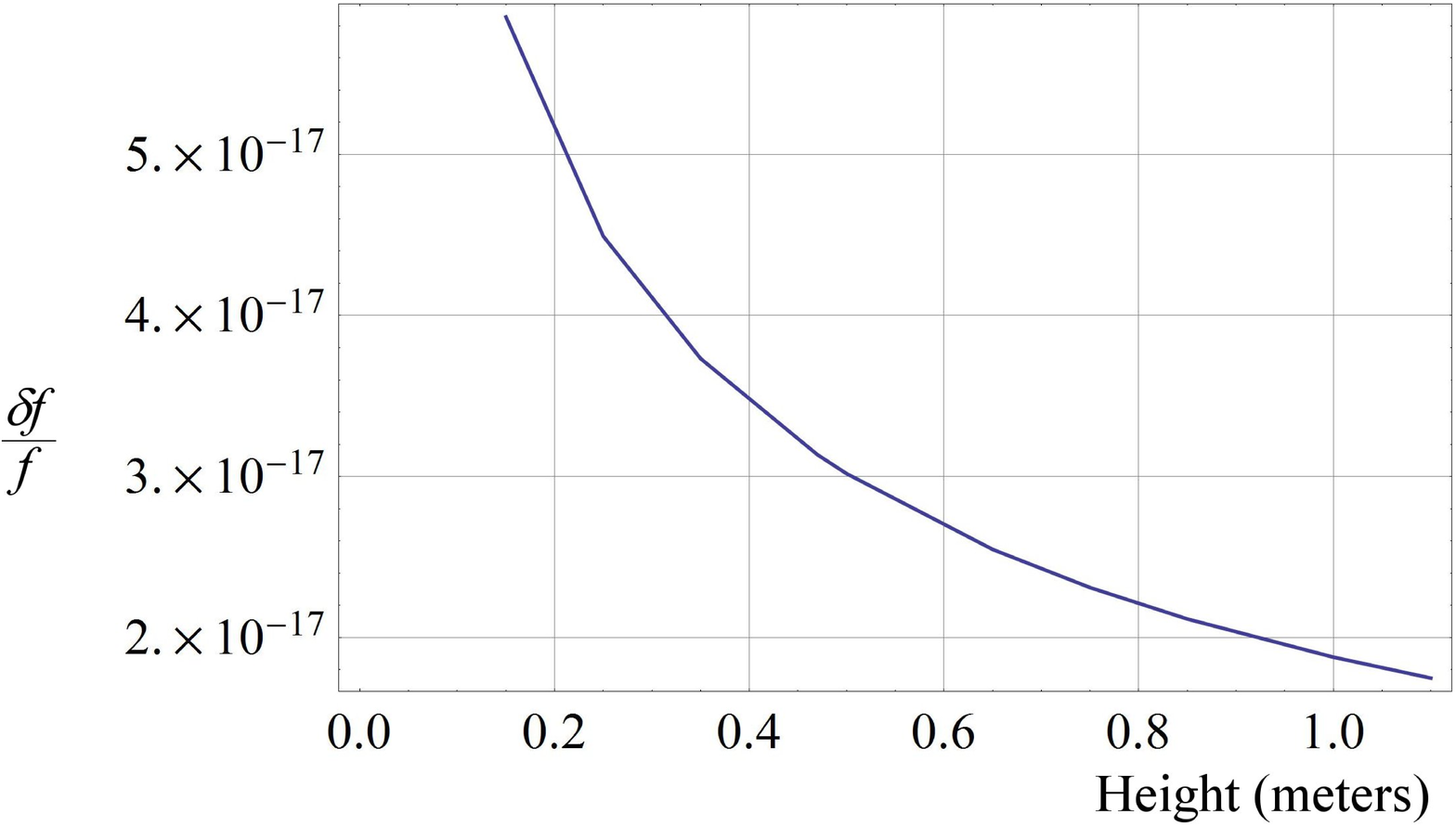}
\caption{Fractional frequency shift vs. toss height for NIST F-2 at temperature $0.5\mu$K; full two-dimensional calculation.}
\end{figure}
Figure 7 plots the fractional frequency shift at a fixed height, as a function of the temperature.  
\begin{figure}
\centering\includegraphics[width=3.5 truein]{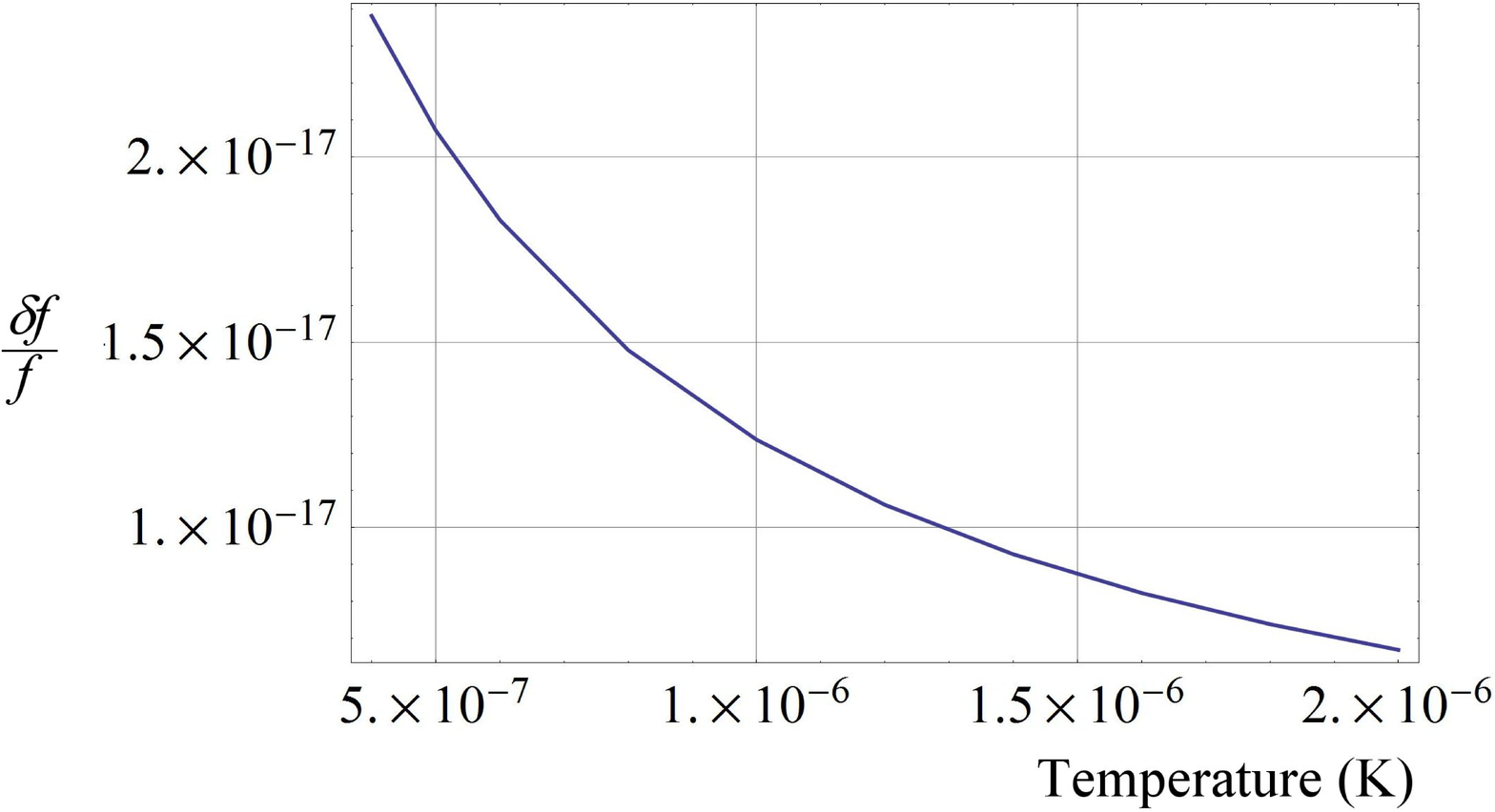}
\caption{Fractional frequency shift for NIST F-2 at a toss height 0.75 meters, as a function of temperature.}
\end{figure}
In Figure 8 we plot the probability of arrival at the detector, of a state-selected and launched atom, as a function of toss height.
The fraction decreases as toss height increases because the atom ball has more time to spread out due to the distribution of thermal velocities.  Figure 9 shows the frequency shift dependence on initial packet width for a toss height of 0.47 m at temperature $0.5\mu$K.
\begin{figure}
\centering\includegraphics[width=3.5truein]{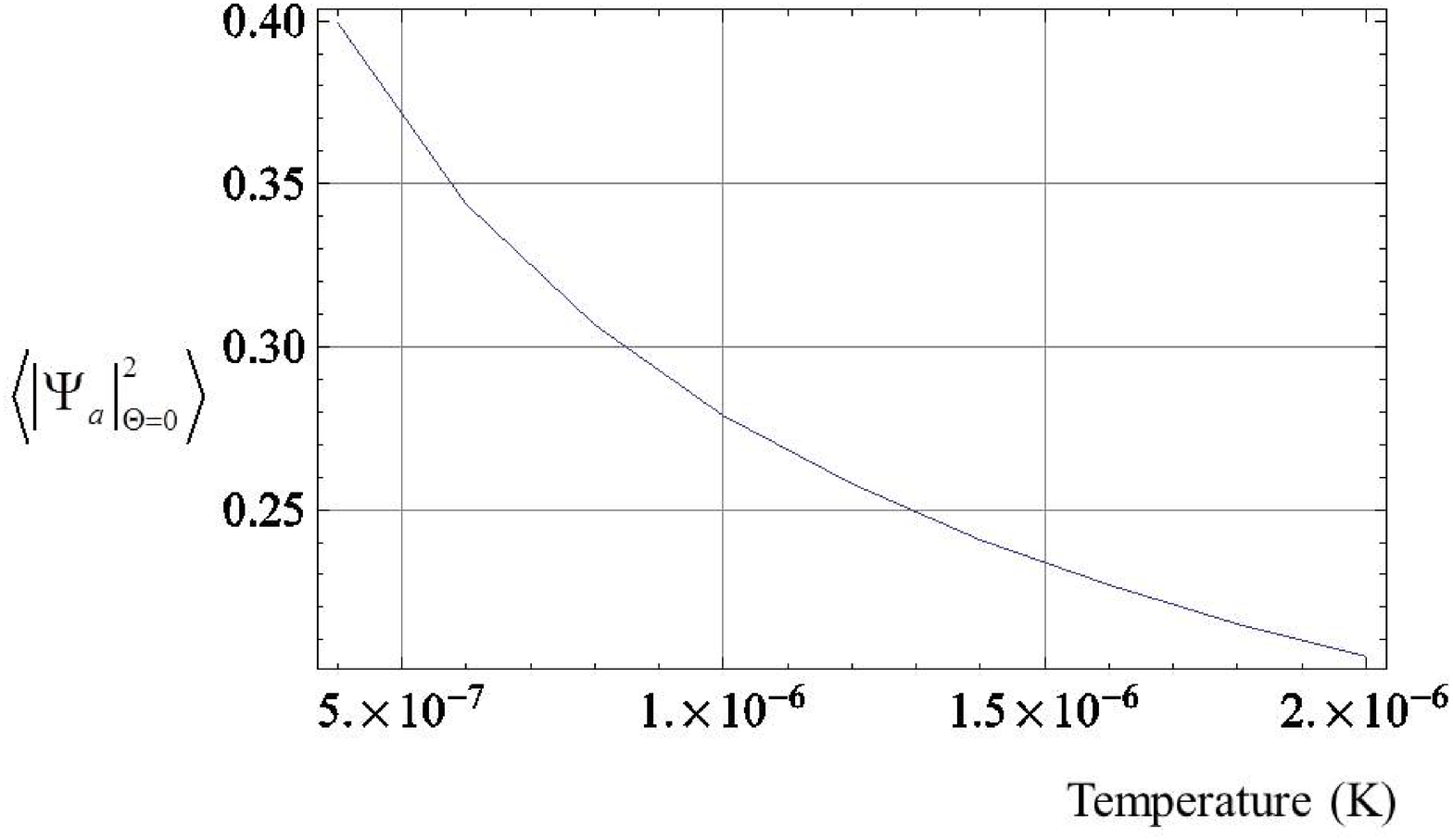}
\caption{Fraction of atoms arriving at the detector as a function of toss height; two dimensional case.}
\end{figure}
The wavepacket widths at launch are unknown; if chosen to be too small, quantum mechanical spreading will become very important and loss of atoms due to clipping by the apertures will become significant.  In Figure 9 we plot the fractional frequency shift as a function of the width parameter $\sigma$.  (The de Broglie wavelength of a Cesium atom at $0.5\mu$K is about 400 nm.)  If the assumed width is larger than about 200 nm, which is the value we have used in all our other calculations, the shift is essentially independent of $\sigma$.  Figure 10 plots the fractional frequency shifts as a function of toss height for NIST F-2 for $\pi/2$, $3\pi/2$, and $5\pi/2$ pulses.
\begin{figure}
\centering\includegraphics[width=3.5truein]{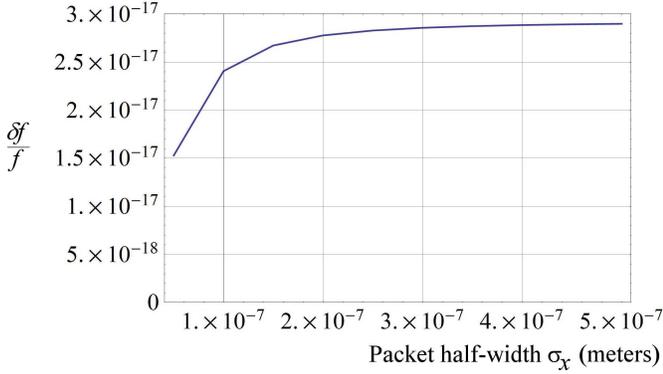}
\caption{Fractional frequency shift for a toss height of 0.47 m and temperature .5$\mu$K as a function of initial packet width.}
\end{figure}
\begin{figure}
\centering\includegraphics[width=3.5truein]{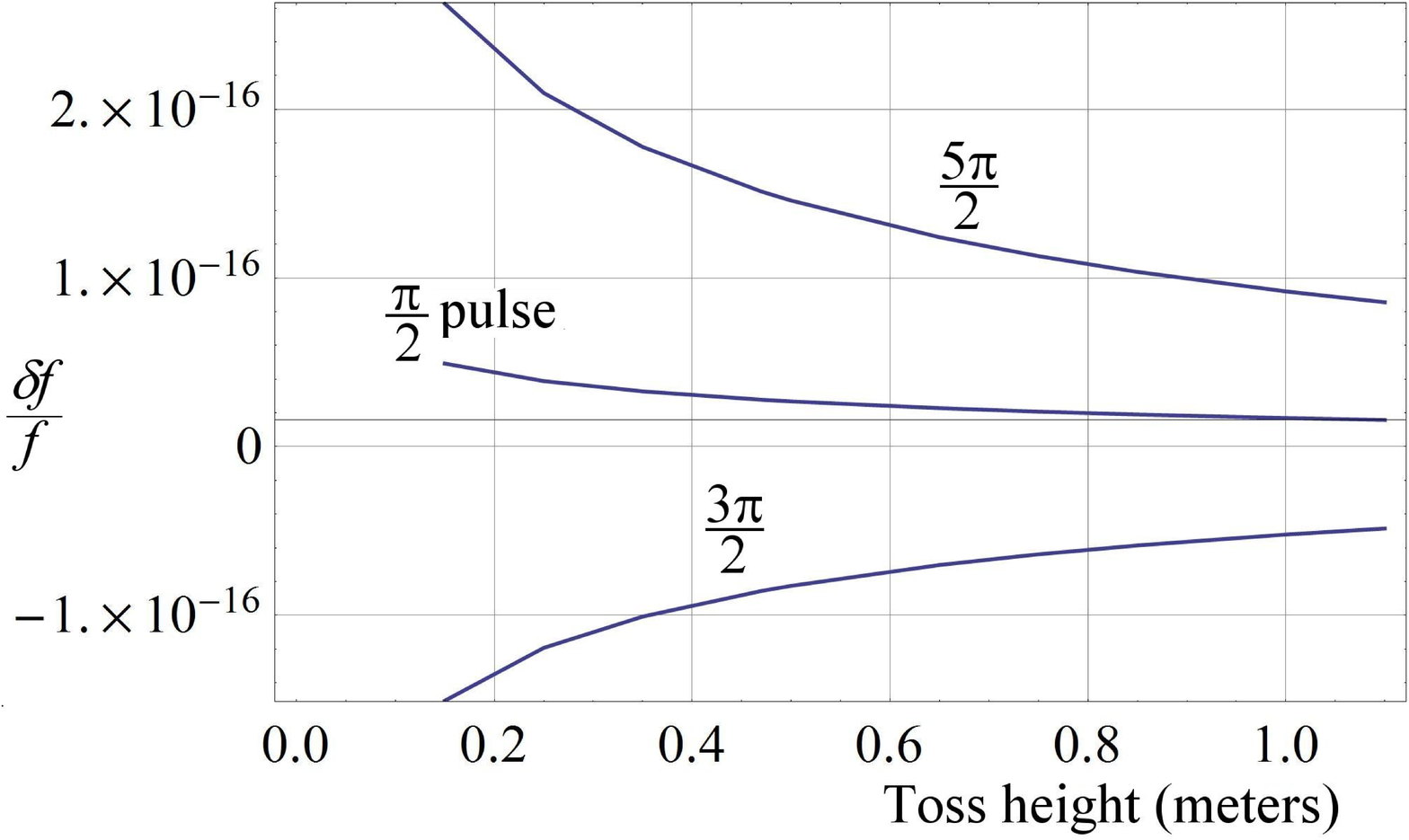}
\caption{Fractional frequency shift as a function of toss height for $\pi/2,\ 3\pi/2,\ 5\pi/2$ pulses.}
\end{figure}

Table I  provides values of the fractional frequency shift as a function of RF amplitude corresponding to $b=\pi/4 \tau,\ 3\pi/4\tau,\ 5\pi/4\tau$, and $7\pi/4\tau$ pulses for different toss heights.  For $b=3\pi/4\tau$ and $7\pi/4\tau$, the $n \pi/2$ the shifts are negative.   The dependence on applied RF amplitude is similar to that of the microwave leakage frequency shift and is accounted for during normal evaluation of the NIST fountains.  The fractional frequency shifts at zero applied rf amplitude, calculated using the method described in this paper, are zero at all launch heights.

\begin{table}
\begin{center}
\caption{Fractional shifts $\times 10^{17}$ for various launch heights as a function of applied microwave field amplitude.  The last line contains the values extrapolated to zero applied rf amplitude.}
\vbox to 8pt {}
\begin{tabular}{l  c  c  c}
\hline
&\multicolumn{3}{c}{height(m)}\\ \hline
\rule[-8pt]{0pt}{22pt}
\quad b\quad &\quad 0.47 \quad & \quad 0.75\quad  & \quad 1.00 \\ \hline
\quad $\pi/4\tau$ & 2.78 & 2.07 & 1.69 \\
\quad $3\pi/4\tau$ & -8.56 & -6.39 & -5.22 \\
\quad $5\pi/4\tau$ & 15.1 & 11.3 & 9.20 \\
\quad $7\pi/4\tau$ & -23.3 & -17.4 & -14.2 \\
\quad 0 (fit)&  -1.2 & -0.9  & -0.7  \\
\hline
\end{tabular}
\end{center}
\end{table}

\section{12. Conclusions}

	For the parameters for NIST F-1 and F-2 Cesium fountains--e.g., cavity radius, configuration of detectors, aperture size, etc., we have not found any conditions such that frequency shifts due to transverse field gradients are large enough to be significant in the systematic error budget.  We have accounted for many factors that make the present calculation realistic.  These include temperature and spatial distributions in the launched atom balls, quantum mechanical spreading of the atomic wave packets, clipping of probability distributions at aperture boundaries, distances between trapping regions, cavities, and detectors that affect times of passage of atoms through the cavities and the time spent in the drift region, transverse motion of atoms in the drift region that results in their sampling different values of the transverse field gradients, off-axis values of the Rabi pedestal, as well as a full two-dimensional theory of atomic trajectories through the cavities based on an exact Green's function solution of the Schr\"odinger equation in the presence of a transverse field gradient.

	If there is a weakness in this approach, it is that use of the Green's function solutions require integrations from $-\infty$ to $+\infty$ in the transverse direction whereas such directions are limited by the cavity apertures.  However wavepackets are concentrated in a very small region and their distributions rapidly approach zero away from their centroids; consequently the contributions to such integrals outside of the apertures are extremely small, but have not been neglected

\section{APPENDIX}
The integral (\ref{eq29}) may be evaluated with the aid of a convergence factor that is allowed to approach zero at the end of the calculation.  Displaying only the factors that participate in the integration, we have
\ba
\lim_{\alpha \to 0}\sqrt{\frac{m}{2\pi i \hbar T}}\int dx'e^{\frac{im(x-x')^2}{2\hbar T}-\alpha^2(x-x')^2+i(k \mp \frac{2\gamma \tau}{\hbar \pi})(x-x')}\nonumber\\
=\lim_{\alpha \to 0}\sqrt{\frac{m}{2\pi i \hbar T}}\sqrt{2\pi}\sqrt{\frac{\hbar T}{-im+2\alpha^2\hbar T}}\hbox to 1in{}\nonumber\\
\times e^{( k \mp \frac{2\gamma \tau}{\hbar \pi})^2/(2im/(\hbar T)-2\alpha^2)}\hbox to 1.3in{}\nonumber\\
=\exp\big( -i\hbar T(k \mp \frac{2\gamma \tau}{\hbar \pi})^2/(2m) \big)\hbox to .5in{} .
\ea
When combined with the other phase factors in (27) this yields (30).
\subsection{}
\subsubsection{}




\bibliography{mybibliography}

\providecommand{\noopsort}[1]{}\providecommand{\singleletter}[1]{#1}%
\begin{thebibliography}{13}%
\makeatletter
\providecommand \@ifxundefined [1]{%
 \@ifx{#1\undefined}
}%
\providecommand \@ifnum [1]{%
 \ifnum #1\expandafter \@firstoftwo
 \else \expandafter \@secondoftwo
 \fi
}%
\providecommand \@ifx [1]{%
 \ifx #1\expandafter \@firstoftwo
 \else \expandafter \@secondoftwo
 \fi
}%
\providecommand \natexlab [1]{#1}%
\providecommand \enquote  [1]{``#1''}%
\providecommand \bibnamefont  [1]{#1}%
\providecommand \bibfnamefont [1]{#1}%
\providecommand \citenamefont [1]{#1}%
\providecommand \href@noop [0]{\@secondoftwo}%
\providecommand \href [0]{\begingroup \@sanitize@url \@href}%
\providecommand \@href[1]{\@@startlink{#1}\@@href}%
\providecommand \@@href[1]{\endgroup#1\@@endlink}%
\providecommand \@sanitize@url [0]{\catcode `\\12\catcode `\$12\catcode
  `\&12\catcode `\#12\catcode `\^12\catcode `\_12\catcode `\%12\relax}%
\providecommand \@@startlink[1]{}%
\providecommand \@@endlink[0]{}%
\providecommand \url  [0]{\begingroup\@sanitize@url \@url }%
\providecommand \@url [1]{\endgroup\@href {#1}{\urlprefix }}%
\providecommand \urlprefix  [0]{URL }%
\providecommand \Eprint [0]{\href }%
\providecommand \doibase [0]{http://dx.doi.org/}%
\providecommand \selectlanguage [0]{\@gobble}%
\providecommand \bibinfo  [0]{\@secondoftwo}%
\providecommand \bibfield  [0]{\@secondoftwo}%
\providecommand \translation [1]{[#1]}%
\providecommand \BibitemOpen [0]{}%
\providecommand \bibitemStop [0]{}%
\providecommand \bibitemNoStop [0]{.\EOS\space}%
\providecommand \EOS [0]{\spacefactor3000\relax}%
\providecommand \BibitemShut  [1]{\csname bibitem#1\endcsname}%
\let\auto@bib@innerbib\@empty
\bibitem [{\citenamefont {Wynands}\ and\ \citenamefont
  {Weyers}(2005)}]{wynands05}%
  \BibitemOpen
  \bibfield  {author} {\bibinfo {author} {\bibfnamefont {R.}~\bibnamefont
  {Wynands}}\ and\ \bibinfo {author} {\bibfnamefont {S.}~\bibnamefont
  {Weyers}},\ }\href@noop {} {\bibfield  {journal} {\bibinfo  {journal}
  {Metrologia}\ }\textbf {\bibinfo {volume} {42}},\ \bibinfo {pages} {S64}
  (\bibinfo {year} {2005})}\BibitemShut {NoStop}%
\bibitem [{\citenamefont {Wolf}\ and\ \citenamefont
  {Bord\'e}(2004)}]{wolfborde2004}%
  \BibitemOpen
  \bibfield  {author} {\bibinfo {author} {\bibfnamefont {P.}~\bibnamefont
  {Wolf}}\ and\ \bibinfo {author} {\bibfnamefont {C.~J.}\ \bibnamefont
  {Bord\'e}},\ }\href@noop {} {\bibfield  {journal} {\bibinfo  {journal}
  {arXiv:quant-ph/0403194}\ } (\bibinfo {year} {2004})}\BibitemShut {NoStop}%
\bibitem [{\citenamefont {Li}\ \emph {et~al.}(2011)\citenamefont {Li},
  \citenamefont {Gibble},\ and\ \citenamefont {Szymaniec}}]{li11}%
  \BibitemOpen
  \bibfield  {author} {\bibinfo {author} {\bibfnamefont {R.}~\bibnamefont
  {Li}}, \bibinfo {author} {\bibfnamefont {K.}~\bibnamefont {Gibble}}, \ and\
  \bibinfo {author} {\bibfnamefont {K.}~\bibnamefont {Szymaniec}},\ }\href@noop
  {} {\bibfield  {journal} {\bibinfo  {journal} {Metrologia}\ }\textbf
  {\bibinfo {volume} {48}},\ \bibinfo {pages} {283} (\bibinfo {year}
  {2011})}\BibitemShut {NoStop}%
\bibitem [{\citenamefont {Gibble}(2006)}]{gibble06}%
  \BibitemOpen
  \bibfield  {author} {\bibinfo {author} {\bibfnamefont {K.}~\bibnamefont
  {Gibble}},\ }\href@noop {} {\bibfield  {journal} {\bibinfo  {journal} {Phys.\
  Rev.\ Letts.}\ }\textbf {\bibinfo {volume} {97}},\ \bibinfo {pages} {073002}
  (\bibinfo {year} {2006})}\BibitemShut {NoStop}%
\bibitem [{\citenamefont {Cook}(1978)}]{cook78}%
  \BibitemOpen
  \bibfield  {author} {\bibinfo {author} {\bibfnamefont {R.~J.}\ \bibnamefont
  {Cook}},\ }\href@noop {} {\bibfield  {journal} {\bibinfo  {journal} {Phys.\
  Rev.\ Letts.}\ }\textbf {\bibinfo {volume} {41}},\ \bibinfo {pages} {1788}
  (\bibinfo {year} {1978})}\BibitemShut {NoStop}%
\bibitem [{\citenamefont {Cook}(1987)}]{cook87}%
  \BibitemOpen
  \bibfield  {author} {\bibinfo {author} {\bibfnamefont {R.~H.}\ \bibnamefont
  {Cook}},\ }\href@noop {} {\bibfield  {journal} {\bibinfo  {journal} {Phys.\
  Rev.\ A}\ }\textbf {\bibinfo {volume} {35}},\ \bibinfo {pages} {3844}
  (\bibinfo {year} {1987})}\BibitemShut {NoStop}%
\bibitem [{\citenamefont {Heavner}\ \emph {et~al.}(2005)\citenamefont
  {Heavner}, \citenamefont {Jefferts}, \citenamefont {Donley}, \citenamefont
  {Shirley},\ and\ \citenamefont {Parker}}]{heavner05}%
  \BibitemOpen
  \bibfield  {author} {\bibinfo {author} {\bibfnamefont {T.~P.}\ \bibnamefont
  {Heavner}}, \bibinfo {author} {\bibfnamefont {S.~R.}\ \bibnamefont
  {Jefferts}}, \bibinfo {author} {\bibfnamefont {E.~A.}\ \bibnamefont
  {Donley}}, \bibinfo {author} {\bibfnamefont {J.~H.}\ \bibnamefont {Shirley}},
  \ and\ \bibinfo {author} {\bibfnamefont {T.~E.}\ \bibnamefont {Parker}},\
  }\href@noop {} {\bibfield  {journal} {\bibinfo  {journal} {Metrologia}\
  }\textbf {\bibinfo {volume} {42}},\ \bibinfo {pages} {411} (\bibinfo {year}
  {2005})}\BibitemShut {NoStop}%
\bibitem [{\citenamefont {Scully}\ \emph {et~al.}(1989)\citenamefont {Scully},
  \citenamefont {Englert},\ and\ \citenamefont {Schwinger}}]{scully89}%
  \BibitemOpen
  \bibfield  {author} {\bibinfo {author} {\bibfnamefont {M.~O.}\ \bibnamefont
  {Scully}}, \bibinfo {author} {\bibfnamefont {B.-O.}\ \bibnamefont {Englert}},
  \ and\ \bibinfo {author} {\bibfnamefont {J.}~\bibnamefont {Schwinger}},\
  }\href@noop {} {\bibfield  {journal} {\bibinfo  {journal} {Phys.\ Rev.\ A}\
  }\textbf {\bibinfo {volume} {40}},\ \bibinfo {pages} {1775} (\bibinfo {year}
  {1989})}\BibitemShut {NoStop}%
\bibitem [{\citenamefont {Shirley}\ \emph {et~al.}(2001)\citenamefont
  {Shirley}, \citenamefont {Lee},\ and\ \citenamefont
  {Drullinger}}]{shirley01}%
  \BibitemOpen
  \bibfield  {author} {\bibinfo {author} {\bibfnamefont {J.~H.}\ \bibnamefont
  {Shirley}}, \bibinfo {author} {\bibfnamefont {W.~D.}\ \bibnamefont {Lee}}, \
  and\ \bibinfo {author} {\bibfnamefont {R.~E.}\ \bibnamefont {Drullinger}},\
  }\href@noop {} {\bibfield  {journal} {\bibinfo  {journal} {Metrologia}\
  }\textbf {\bibinfo {volume} {38}},\ \bibinfo {pages} {427} (\bibinfo {year}
  {2001})}\BibitemShut {NoStop}%
\bibitem [{\citenamefont {Schwinger}\ \emph {et~al.}(1988)\citenamefont
  {Schwinger}, \citenamefont {Scully},\ and\ \citenamefont
  {Englert}}]{schwinger88}%
  \BibitemOpen
  \bibfield  {author} {\bibinfo {author} {\bibfnamefont {J.}~\bibnamefont
  {Schwinger}}, \bibinfo {author} {\bibfnamefont {M.~O.}\ \bibnamefont
  {Scully}}, \ and\ \bibinfo {author} {\bibfnamefont {B.-G.}\ \bibnamefont
  {Englert}},\ }\href@noop {} {\bibfield  {journal} {\bibinfo  {journal}
  {Zeits.\ Phys.\ D}\ }\textbf {\bibinfo {volume} {10}},\ \bibinfo {pages}
  {135} (\bibinfo {year} {1988})}\BibitemShut {NoStop}%
\bibitem [{\citenamefont {Englert}\ \emph {et~al.}(1988)\citenamefont
  {Englert}, \citenamefont {Schwinger},\ and\ \citenamefont
  {Scully}}]{englert88}%
  \BibitemOpen
  \bibfield  {author} {\bibinfo {author} {\bibfnamefont {B.-G.}\ \bibnamefont
  {Englert}}, \bibinfo {author} {\bibfnamefont {J.}~\bibnamefont {Schwinger}},
  \ and\ \bibinfo {author} {\bibfnamefont {M.~O.}\ \bibnamefont {Scully}},\
  }\href@noop {} {\bibfield  {journal} {\bibinfo  {journal} {Found.\ Phys.}\
  }\textbf {\bibinfo {volume} {18}},\ \bibinfo {pages} {1045} (\bibinfo {year}
  {1988})}\BibitemShut {NoStop}%
\bibitem [{\citenamefont {Shulman}(1981)}]{shulman65}%
  \BibitemOpen
  \bibfield  {author} {\bibinfo {author} {\bibfnamefont {L.~S.}\ \bibnamefont
  {Shulman}},\ }\href@noop {} {\emph {\bibinfo {title} {Techniques and
  Applications of Path-Integral Methods}}}\ (\bibinfo  {publisher} {Wiley},\
  \bibinfo {year} {1981})\ p.~\bibinfo {pages} {38}\BibitemShut {NoStop}%
\bibitem [{\citenamefont {Kennard}(1927)}]{Kennard27}%
  \BibitemOpen
  \bibfield  {author} {\bibinfo {author} {\bibfnamefont {E.~H.}\ \bibnamefont
  {Kennard}},\ }\href@noop {} {\bibfield  {journal} {\bibinfo  {journal}
  {Zeits.\ Phys.}\ }\textbf {\bibinfo {volume} {44}},\ \bibinfo {pages} {326}
  (\bibinfo {year} {1927})}\BibitemShut {NoStop}%
\end{thebibliography}%

\end{document}